\documentclass[12pt]{article}
\usepackage[space]{grffile}
\usepackage{latexsym}
\usepackage{pgfplots}
\pgfplotsset{compat=1.17}
\usepackage{textcomp}
\usepackage{longtable}
\usepackage{tabulary}
\usepackage{booktabs,array,multirow}
\usepackage{amsfonts,amsmath,amssymb}
\usepackage{natbib}
\usepackage{url}
\usepackage{hyperref}
\hypersetup{colorlinks=false,pdfborder={0 0 0}}
\usepackage{etoolbox}
\makeatletter
\usepackage{amsmath,amsfonts,bbm,xfrac}
\usepackage{graphicx,float}
\usepackage{hyperref}
\usepackage{xcolor}
\usepackage{color,soul}
\usepackage{graphicx}
\graphicspath{ {./figs/} }
\usepackage[english]{babel}
\newtheorem{theorem}{Theorem}[section]

\usepackage{algorithm}
\usepackage{algorithmic} 
\usepackage{authblk}

\newcommand{\R}{\mathbb{R}}

\newcommand{\E}{\mathbb{E}}

\newcommand{\paren}[1]{{\left(#1\right)}} 
\newcommand{\brac}[1]{{\left\{#1\right\}}} 
\newcommand{\braces}[1]{{\left[#1\right]}} 
\newcommand{\norm}[1]{{\left\|#1\right\|}} 

\makeatother
\newif\iflatexml\latexmlfalse

\AtBeginDocument{\DeclareGraphicsExtensions{.pdf,.PDF,.eps,.EPS,.png,.PNG,.tif,.TIF,.jpg,.JPG,.jpeg,.JPEG}}

\usepackage[utf8]{inputenc}
\usepackage[english]{babel}
\newcommand{\blind}{0}

\begin{document}

\def\spacingset#1{\renewcommand{\baselinestretch}%
{#1}\small\normalsize} \spacingset{1}


\if0\blind 
{
  \title{\bf Corrected Correlation Estimates for Meta-Analysis}
    \author[1,2]{Alexander Johnson-Vázquez\thanks{
    The authors gratefully acknowledge the Bill \& Melinda Gates Foundation.}}
    \author[1,2]{Alexander W. Hsu}
    \author[1,2]{Aleksandr Aravkin}
    \author[1]{Peng Zheng}

    \affil[1]{Institute for Health Metrics and Evaluation, Seattle}
    \affil[2]{Department of Applied Mathematics, University of Washington, Seattle}

  \maketitle
} \fi

\if1\blind
{
  \bigskip
  \bigskip
  \bigskip
  \begin{center}
    {\bf Corrected Correlation Estimates for Meta-Analysis}
\end{center}
  \medskip
} \fi

\bigskip
\begin{abstract}
Meta-analysis aggregates estimates and uncertainty across multiple studies, summarizing individual reports into aggregate results that are frequently used to inform health policy and recommendations. When a given study reports multiple estimates, such as log odds ratios (ORs) or log relative risks (RRs) across different exposure groups, accounting for within-study estimate correlations is improves both efficiency of meta-analytic estimates and provides more accurate estimates of uncertainty. The canonical approaches of  \cite{Greenland1992MethodsFT} and \cite{hamling2008facilitating} construct pseudo-cases and non-cases for exposure groups to estimate correlations of reported within-study estimates. However,  currently availble implementations for both methods can fail on simple examples.

We review both GL and Hamling methods through the lens of optimization. For ORs, we provide modifications of each approach that ensure convergence for any feasible inputs. For GL, this is achieved through a new connection to entropic minimization. For Hamling, a modification leads to a provably solvable equivalent set of equations given a specific initialization. For each, we provide  implementations
guaranteed to work for any feasible input. 

For RRs, we show the new GL approach is always guaranteed to succeed. We derive counter-examples where the Hamling approach does not admit any solutions. For the special RR case where the variances are all equal, we derive a necessary and sufficient condition for success.
\end{abstract}

\noindent%

\section{Introduction}\label{section:1}
Meta-analysis combines results reported by multiple studies to obtain aggregate results and estimate between-study heterogeneity~\cite{haidich2010meta}. 
Meta-analytic results inform public health recommendations, underscoring the importance of accuracy in meta-analytic methods~\citep{deeks2019analysing}[Chapter 10]. 
Understanding dose-response relationships across different ranges of exposure poses particular challenges~\cite{orsini2012meta,liu2009two,crippa2019one,zheng2022burden}. 
Dose-response meta-analysis seeks to quantify the impact of a continuous risk, such as
systolic blood pressure~\citep{razo2022effects}, 
smoking~\citep{dai2022health}, 
meat~\citep{lescinsky2022health} or vegetables~\citep{stanaway2022health} 
consumed, 
on the risk of an outcome, e.g. lung cancer or heart disease, by aggregating available estimates for different exposure groups across many studies. 

Two of the most common types of estimates are adjusted odds ratios and relative risks~\citep{schmidt2008use}. Because these estimates always share a common reference group, the estimates for different exposure levels are correlated. Estimating relationships without correcting for these correlations is inefficient and under-estimates the variance of the resulting coefficients~\cite[ Appendix (1)]{Greenland1992MethodsFT}. We show the potential impact of the adjustment, as well as a real-world example, in Section~\ref{sec:motivation}.

In short, it is crucial for meta-analyses to adjust for within-study correlation. Since we are blind to the adjustment mechanism of reported odds ratios (ORs) and relative risks (RRs), we do not have access to the true underlying covariance matrix between reported estimates. If the adjusted estimates are produced through a regression, then an estimated covariance matrix would be available. However, this estimated covariance matrix is generally not reported. As we have access only to the reported metadata, we must accurately construct this covariance matrix.
In their groundbreaking work, \cite{Greenland1992MethodsFT} showed that it is possible to estimate 
within-study correlations, and use them to approximate the covariance matrix. 
The GL approach  requires the modeler to provide the total number of subjects at each exposure level (both treatment and control), the total number of cases, and adjusted treatment effects at each exposure level, such as log ORs or log RRs.  
Using this information, the GL approach uses a root-finding algorithm to obtain pseudo-case counts for every exposure that match reported estimates, 
and then uses the pseudo-counts to estimate asymptotic within-study correlations. 
These correlations inform downstream analyses, accounting for the impact of a common reference group explicitly before estimating study-specific random effects through mixed-effects modeling. 

Following the work of~\cite{Greenland1992MethodsFT}, \cite{hamling2008facilitating} also use reported estimates to get pseudo-counts of cases versus non-cases. However, \cite{hamling2008facilitating} directly use the standard errors of the reported estimates rather than requiring modelers to obtain
subject counts at each exposure level. 
The Hamling approach  requires only two additional pieces of information beyond the estimates and their variances: the ratio of unexposed controls to total exposed controls, and the ratio of all controls to all cases.  
\cite{hamling2008facilitating} fit pseudo-cell counts to the available data, and given pseudo-cell counts, the correlation estimators are the same as those of GL. 

These methods are widely used in the community; for example, the meta-analysis \verb+R+ package \verb+dosresmeta+ \citep{dosresmeta} implements both correlation estimators 
in their \verb+Covariance+ function 
that creates the within-study covariance matrix. Despite the wide use of both methods, past research stopped short of providing  guarantees of success given feasible inputs. In fact, both~\cite{Greenland1992MethodsFT} and~\cite{hamling2008facilitating} discussed numerical instability, citing occasional failures and the need to re-initialize as needed. As originally presented, and as currently implemented in~\cite{dosresmeta}, both methods fail on simple modifications to the input data from working examples. 

Here, we fill the current gap, providing robust GL and Hamling methods guaranteed to work for all feasible inputs on the OR problem, including our generated failure modes that can break the current implementation~\cite{dosresmeta}. To do this, we study each approach using an optimization perspective.  
For GL, we show the root-finding problem of~\cite{Greenland1992MethodsFT} is equivalent to a convex minimization problem in both the OR and RR settings.
Convexity allows us to prove existence and uniqueness of results, and use disciplined convex programming (DCP)~\citep{boyd2004convex} to remove any decisions by the user regarding initialization and to provide state-of-the-art numerical solving techniques. We provide an implementation using \verb|cvxpy| that is guaranteed to return the unique solution~\citep{diamond2016cvxpy, agrawal2018rewriting}.
For Hamling, in the case of OR, we develop an  equivalent set of nonlinear equations, and prove that these equivalent formulations are always solvable.  
We provide a  Python implementation that, in practice, converges for all inputs. For RR, we show that in fact the Hamling approach may fail, provide a counter-example where there is no solution, and provide a sufficient condition on solvability for reported RR's 
in the case where reported variances are all equal. 
Our implementation also covers the RR case but provides an informative warning to the modeler should the model fail to find a root. 

\paragraph{Roadmap.} In Section~\ref{sec:motivation}, we provide theoretical and empirical motivation for 
adjusting for within-study correlation, which may be  useful for readers new to the topic. We  review the work of \cite{Greenland1992MethodsFT} and \cite{hamling2008facilitating} in Section~\ref{sec:GLHam}. We develop the necessary innovations to robustify each method and provide theoretical guarantees in Sections~\ref{Cvx:GL} and~\ref{sec:solve}.  
Finally, in Section~\ref{sec:numerics}, we present numerical illustrations showing our methods provide identical results to those of~\cite{Greenland1992MethodsFT} and~\cite{hamling2008facilitating} when the original methods converge, and provide correct results for inputs that break currently available implementation. 
We also present a counter-example in the RR regime  that has no solution for the Hamling approach. 

\section{Motivation for Correlation Correction}
\label{sec:motivation}

Before we review existing methods and introduce our updated techniques for correcting for within-study correlation, we motivate the necessity of such methods. We show that considering differences in means with respect to a reference group always induces a nonzero correlation reported estimates. Building on this example, we construct a toy simulation that shows the potential impact of failing to account for this correlation. We show a simple example from a real-world study in peripheral artery disease in which we observe high correlation between estimates, leading to a significant difference in the slope of the dose-response relationship between adjusted and unadjusted estimates.
Finally, we briefly describe implications of the correlation correction for meta-analysis. 


\subsection{Theoretical Motivation}
Consider measurements $\brac{x_1^i}_{i=1}^{n_1}$,$\brac{x_2^j}_{j=1}^{n_2}$ from two different treatment groups 
and measurements from a reference group,$\brac{x_0^l}_{l=1}^{n_0}$,
where $n_k$ is the number of samples in group $k\in\{0,1,2\}$. Here, we assume that each $x_k^i$ is independently distributed according to a Gaussian distribution distinct for each $k$, i.e., 
\[
    x_k^i \overset{\mathrm{iid}}{\sim} \mathcal{N}(\mu_k, \sigma_k^2)
\]
for nonzero $\mu_k, \sigma_k$. Without loss of generality, assume $\hat\mu_2 > \hat\mu_0$ and $\hat\mu_1 > \hat\mu_0$.

We define the empirical mean estimator as 
\[
    \hat \mu_k = \frac{1}{n_k}\sum_{i=1}^{n_k} x_k^i
\]
and seek to estimate the difference in means between the treatment groups and reference group, constructing the estimators $\hat{\eta}_1 = (\hat \mu_1 - \hat \mu_0)$ and $\hat{\eta}_2 = (\hat \mu_2 - \hat \mu_0)$. The reference group induces a positive correlation between these estimators, as shown below: 
\begin{align*}
    \mathrm{Cov}(\hat{\eta}_1,\hat{\eta}_2) &= \E\braces{\hat{\eta}_1\hat{\eta}_2} - \E\braces{\hat{\eta}_1}\E\braces{\hat{\eta}_2}\\
    &= \E\braces{(\hat \mu_1 - \hat \mu_0)(\hat \mu_2 - \hat \mu_0)} - \E\braces{(\hat \mu_1 - \hat \mu_0)}\E\braces{(\hat \mu_2 - \hat \mu_0)}\\
    &= \E\braces{\hat\mu_1 \hat\mu_2 - \hat\mu_1 \hat\mu_0 - \hat\mu_2 \hat\mu_0 + \hat\mu_0^2} - (\mu_1 - \mu_0)(\mu_2 - \mu_0)\\
    &= \E\braces{\hat\mu_1 \hat\mu_2} - \E\braces{\hat\mu_1 \hat\mu_0} - \E\braces{\hat\mu_2 \hat\mu_0} + \E\braces{\hat \mu_0^2} - \paren{\mu_1\mu_2 - \mu_1\mu_0 - \mu_2\mu_0 + \mu_0^2}\\
    &=\E\braces{\hat\mu_0^2} - \mu_0^2\\
    &= \sigma_0^2/n_0.
\end{align*}

The correlation is driven by the variance of the mean of the reference group. 
By independence, the variance of the estimators themselves is given by 
\[
\mathbb{V}[\hat{\eta}_1] = \frac{\sigma_0^2}{n_0} + \frac{\sigma_1^2}{n_1}, \quad 
\mathbb{V}[\hat{\eta}_2] = \frac{\sigma_0^2}{n_0} + \frac{\sigma_2^2}{n_2}
\]
where $\mathbb{V}$ is the variance operator. Thus, the smaller the reference group, and the larger its intrinsic variance, the larger the induced correlation between $\hat{\eta}_1$ and $\hat{\eta}_2$. 

Using this data and assuming there is a true, linear effect across groups $\beta$, we may seek to estimate $\beta$ through least-squares regression. The two methods we observe are generalized least squares (GLS) and ordinary least squares (OLS). We set $X$ to be the appended vector $X = \paren{\brac{x_0^l}_{l=1}^{n_0}, \brac{x_1^i}_{i=1}^{n_1}, \brac{x_2^k}_{k=1}^{n_2}}^\top$ and also set $\hat\eta = \paren{\hat\eta_1,\hat\eta_2}$. Thus, we may construct the estimator $\hat\beta_{\mathrm{cor}}$ for $\beta$ to be
\[
    \hat \beta_{\mathrm{cor}} = \paren{X^\top C^{-1} X}^{-1}X^\top C^{-1}\hat{\eta}
\]
to be the GLS estimate, where we are accounting for correlation between $\hat\eta_1, \hat\eta_2$, which we know must exist~\citep{kariya2004generalized}. Here, $C$ is the covariance matrix of the estimates $\hat\eta_1,\hat\eta_2$ with entries defined as above. Similarly, we construct the estimate $\hat \beta_{\mathrm{OLS}}$ to $\beta$ to be
\[
    \hat \beta_{\mathrm{OLS}} = \paren{X^\top X}^{-1}X^\top\hat{\eta}.
\]
Note that this OLS estimator does not account for correlation and amounts to assuming the independence of $\hat\eta_1,\hat\eta_2$.

In evaluating these two estimators, it is easy to show that both $\hat \beta_{\mathrm{cor}}, \hat \beta_{\mathrm{OLS}}$ are unbiased. By construction, we have
\begin{align*}
    \mathbb{V}\braces{\hat \beta_{\mathrm{cor}}} &= \paren{X^\top C^{-1}X}^{-1} \\
    \mathbb{V}\braces{\hat \beta_{\mathrm{OLS}}} &= \paren{X^\top X}^{-1}X^\top C X\paren{X^\top X}^{-1}
\end{align*}
as the variance estimators. From the generalized Gauss-Markov theorem~\citep{kariya2004generalized}, it follows that $\hat \beta_{\mathrm{cor}}$ is optimal among all linear, unbiased estimators and asymptotically efficient. In particular, 
$\mathbb{V}\braces{\hat \beta_{\mathrm{cor}}} \leq \mathbb{V}\braces{\hat \beta_{\mathrm{OLS}}}$, with strict inequality whenever $C$ is not diagonal.
This explains the advantage of GLS estimation according to problems of this class. A theme of the present work is that the covariance matrix $C$ is not always known. This further illustrates the necessity of developing good approximation techniques to $C$ so that estimators downstream remain more precise.

In the next section we illustrate the impact of this correlation on the efficiency of the estimator for the overall 
relationship computed from multiple reported estimates. In the context of meta regression, we would often consider both of the exposure effect estimates $\hat{\eta}_1$, $\hat{\eta}_2$ in conjunction with data from other studies to estimate effects as a function of exposure level.


\subsection{Numerical Illustration}
\begin{figure}[h!]
  \centering
  \includegraphics[width=0.8\textwidth]{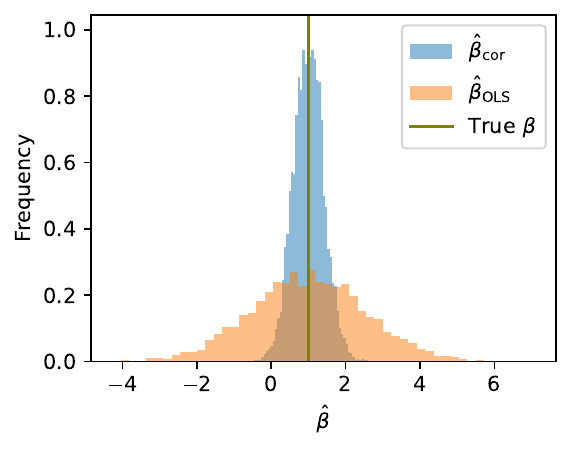} 
  \caption{The empirical distribution of the $\hat \beta$ estimate in a numerical simulation when estimated using GLS with a weighting covariance matrix (blue; $\hat \beta_{\mathrm{cor}}$) versus when estimated using OLS, assuming no correlation (orange; $\hat \beta_{\mathrm{OLS}}$). The true value of $\beta = 1$ prescribed before the simulation is given by the vertical line.
  }
  \label{fig:dist_plot}
\end{figure}

Here, we create a simplified simulation to show the impact of adjusting for the correlation between the mean estimator levels $\hat\eta_i$. In our simulation, using the notation from above, we have $n_0 = 1$ with 4 exposure levels and $n_1 = \dots = n_{4} = 10$. We prescribe a true value of $\beta = 1$ and seek to estimate this population value according to the data.

After assigning all the initial count data, we construct our estimates $\hat\eta_1, \dots, \hat\eta_4$. Using the exposure levels as the standard exogenous variable in the regression, we then have all the relevant data. We compare the estimates $\hat \beta_{\mathrm{cor}}$ and $\hat \beta_{\mathrm{OLS}}$ to $\beta$ using the GLS and OLS formulations as constructed above, with the relevant dimensional differences applied. The results from 5,000 realizations are shown in Figure~\ref{fig:dist_plot}. Both estimators are unbiased, but $\hat \beta_{\mathrm{cor}}$ is has a much smaller variance than $\hat \beta_{\mathrm{OLS}}$. 


The simulation is relevant to the situations  considered by \cite{Greenland1992MethodsFT} and \cite{hamling2008facilitating}, since $\log$ OR and RR estimates are created with respect to the same reference group per study. 
The main differences are that we have explicit access to the full covariance between our reported estimators, while in meta-analytic settings the correlation is hidden and must be inferred--this is the core problem that correlation correction estimators seek to solve.  

We end the section with a real-world example where the adjustment makes a big difference to summarizing the study. 

\subsection{Real example: blood pressure and peripheral artery disease.}
\begin{figure}[h!]
  \centering
  \includegraphics[width=0.8\textwidth]{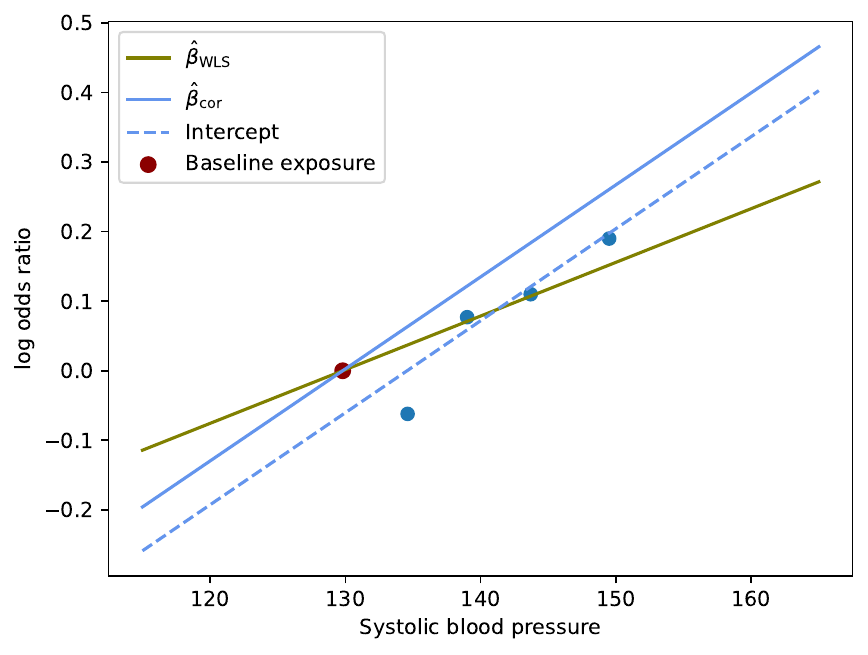} 
  \caption{The regression lines generated by the slope coefficients $\hat \beta_{\mathrm{cor}}$ and $\hat \beta_{\mathrm{WLS}}$ (solid lines). The dashed line is produced by the slope $\hat \beta_{\mathrm{cor}}$ whose intercept is allowed to be non-zero. The baseline exposure is the reference level of exposure considered by the study, used to create relevant OR estimates.}
  \label{fig:itoga_plot}
\end{figure}

We provide a brief example of a real-world study where the correlation-adjusted estimates are significantly different from the estimates obtained when independence of the estimates are assumed.
\cite{itoga2018association} study the impact of blood pressure on peripheral artery disease (PAD), reporting results by subgroups of exposure. We assume a linear relationship between SBP and relative risk of PAD, and visualize the weighted least squares (WLS) and correlation-corrected GLS regressions in Figure~\ref{fig:itoga_plot}. In the meta-analytic setting, studies typically report standard errors. We compare to a naive WLS estimator $\hat \beta_{\mathrm{WLS}}$ where residuals are weighted by the inverse of the reported standard errors. Mathematically, setting $V$ to be the diagonal matrix whose diagonal elements are the reported variances for each exposure level, we have the following estimator:
\[
    \hat\beta_{\mathrm{WLS}} = \paren{X^\top V^{-1} X}^{-1}X^\top V^{-1}\hat{\eta},
\]
where, in this case, $\hat{\eta}$ is the vector of log ORs for each exposure. Using the metadata, we do not have access to the true covariance matrix $C$; we estimate the covariance matrix used in GLS by the method of ~\cite{Greenland1992MethodsFT}. 

 The $x$-axis shows SBP, while the $y$-axis gives the log relative risk. The blue dots show reported adjusted odds ratios, plotted at the mid-points of the exposure groups reported by the paper. 
The solid lines correspond to $\hat \beta_{\mathrm{cor}}$ (blue) and $\hat \beta_{\mathrm{WLS}}$ (olive) in Figure~\ref{fig:itoga_plot}, which are required to pass through the red `origin' point corresponding to the reference group with midpoint at SBP. 
The WLS estimate $\hat \beta_{\mathrm{WLS}}$ appears to fit the data more closely than $\hat \beta_{\mathrm{cor}}$. However, the line produced by $\hat \beta_{\mathrm{cor}}$ provides a better estimate for the slope. 
We can think of it as `adjusting' for the fact that variance in the reference group results propagates to all non-reference points, shifting them up and down together. We illustrate this by including the dashed line in Figure~\ref{fig:itoga_plot}. This is the correlation-corrected estimate shifted to the non-reference data--it should capture the trend of the data more accurately than the WLS estimate. 

In the case of SBP vs. PAD, adjusting for within-study correlation would give a higher estimate of overall risk for that study. While it is impossible to know `truth' for any given study, the simulation in Figure~\ref{fig:dist_plot} serves as a reminder that although both the WLS and GLS estimates are unbiased, the WLS has much higher variance.

We proceed to consider the case of meta-analysis where multiple studies are observed and discuss implications of correcting for correlation in that setting.

\subsection{Implications for Meta-Analysis}

A general description of likelihood formulations for meta-analysis is developed by~\cite{zheng2021trimmed}. 
Taking the simplest example, consider the statistical model for aggregating
multiple reported result vectors $\eta_i$ with specific effects for study $i$: 
\[
\hat \eta_i = X_i \beta + \mathbf{1} u_i + \epsilon_i, 
\]
where $\epsilon_i \sim \mathcal{N}(0, V_i)$ describes the observation errors for study $i$, while $u_i$ is a scalar realization of a random effect distributed as $\mathcal{N}(0, \gamma)$ where $\gamma$ 
represents between-study heterogeneity. 
The $\epsilon_i$ and $u_i$ are independent across $i$, and also from each other. 
This model applies the realization of the specific effect to all observations from study $i$, hence the vector $\textbf{1}$ that copies $u_i$ to impact every element of $\hat \eta_i$. The variance of the error term is given by:
\[
\mathbb{V}[\mathbf{1}u_i + \epsilon_i] = V_i + \gamma \mathbf{1}\mathbf{1}^T. 
\]
The maximum likelihood estimate for $\beta$ and $\gamma$
is then given by solving  
\[
\min_{\beta, \gamma} \sum_i (X_i \beta - \hat \eta_i)^T
\left(V_i + \gamma \mathbf{1}\mathbf{1}^T\right)(X_i \beta - \hat \eta_i) + \frac{1}{2}\log |V_i + \gamma \mathbf{1}\mathbf{1}^T|
\]

From the likelihood expression, we can observe that meta-analysis effectively quantifies the extent to which the reported variances $V_i$ do not represent the inherent variance in the data, and adjust through augmenting with the between-study heterogeneity variance $\gamma$. 
Using only the reported variances corresponds to assuming that each reported $V_i$ is diagonal, so any 
correlation is left to meta-analysis to discover, with a single parameter $\gamma$. In fact, as shown in the previous sections, within-study correlations are induced by the shared reference group, and the extent this happens can vary by study (for example, a study with a very large reference group will have less correlation than a study with a small reference group). As a result, providing correlated $V_i$ will leave $\gamma$ to capture the variances of the unknown study-specific effects, exactly as intended, rather than trying to capture all the residual correlations.  

With the motivation established, we proceed to review methods that are actually used to estimate and compute the correlation used for the correction in this section. For the remainder of the paper, we focus on reliability and accuracy of the correlation correction methods.

\section{Methods of GL and Hamling}
\label{sec:GLHam}
In this section, we present the approaches of GL and Hamling.
In this review section, we focus on 
log ORs to vastly simplify presentation; however our robust methods in Sections~\ref{Cvx:GL} and~\ref{sec:solve} cover both log ORs and log RRs. 
Special challenges and counter-examples for the Hamling approach in the RR case are also presented in Section~\ref{sec:solve}. 

We start by defining key variables following original notation, see Table~\ref{table:notation}. 
\begin{table}[h!]
\caption{Notation and method requirements table\label{table:notation}. }
\begin{center}
\begin{tabular}{|c|c|c|c|} \hline
Variable & Dimension & Definition & Used by  \\ \hline 
$n$ & $1$ & number of alternative exposure levels & -\\ 
\hline
$x$ & $n$ & alternative exposure levels & -\\
\hline
$N$ & $n+1$ & total subjects at all exposures & GL\\
\hline
$M_1$ & $1$ & total cases & GL\\
\hline 
$L$ &  $n$ & estimates of log-odds & GL, H \\
\hline 
$V$ & $n$ & reported variances for log-odds & H \\ 
\hline 
$R$ &  $n$ & estimates of log-risks & GL, H\\
\hline 
$V^R$ & $n$ & reported variances for log-risks & H \\ 
\hline 
$A$ & $n$ & cases for alternative exposures & - \\
\hline
$a_0$ &$1$ & cases for reference exposure & -\\ 
\hline
$B$ & $n$ & non-cases for alternative exposures & -\\
\hline
$b_0$ & $1$ & non-cases for reference exposure & -\\
\hline 
$p$ & $1$ & ratio of unexposed controls to total  controls & H\\
\hline
$z$ & $1$ & ratio of total controls to total cases & H\\
\hline
\end{tabular}
\end{center}
\end{table}
$M_1$ is the sum of all elements of $A$ and $a_0$. For both GL and Hamling,  
the goal is to estimate $A, a_0, B$, and $b_0$. Following~\cite{Greenland1992MethodsFT} and~\cite{hamling2008facilitating}, we refer to the first element in the vector $N$ as $n_0$ and the remaining elements as $N_+$. We always have that $A + B = N_+$ and $a_0 + b_0 = n_0$. We also include the data requirements by each study. More details are given in the following sections for how the data are used. Here, H is shorthand for the Hamling method.
With notation established, we summarize the main goal of the GL and Hamling methods.

\subsection{Correlation and Covariance}\label{sec:cor_cov}

The main goal of both GL and Hamling methods is to obtain a variance-covariance matrix, replacing a diagonal matrix of reported variances with an updated variance-covariance matrix with the same variances and estimated correlations. In particular, both methods estimate the correlation for two log ORs at two different exposures $x_i$ and $x_j$ by 
\begin{equation}
\label{eq:corrOR}
r_{x_i, x_j} = \frac{1/a_0 + 1/b_0}{\sqrt{1/a_0 + 1/b_0 + 1/A_i + 1/B_i}\sqrt{1/a_0 + 1/b_0 + 1/A_j + 1/B_j}} 
\end{equation}
where $B_i$ represent controls, 
and the correlation for two log RRs at these exposures by 
\begin{equation}
\label{eq:corrRR}
r_{x_i, x_j} = \frac{1/a_0 - 1/b_0}{\sqrt{1/a_0 - 1/b_0 + 1/A_i - 1/B_i}\sqrt{1/a_0 - 1/b_0 + 1/A_j - 1/B_j}} 
\end{equation}
where $B_i$ represent totals. The final variance-covariance matrix is obtained by appropriately scaling these correlations using the reported variances. There is a degree of freedom in the pseudo-counts that factors out of the correlation formulas: all pseudo-counts can be multiplied by a constant value and the correlations would not change in either the OR or the RR case.   
 
Finally it may help to alert the reader to the key difference between the Hamling and GL approaches by observing that by construction of the Hamling approach, the pseudo-counts successfully obtained by that method (for either RRs or ORs) satisfy
\[
r_{x_i, x_j} = \frac{1/a_0 + 1/b_0}{\sqrt{V_iV_j}}
\]
where $V_i$, $V_j$ are the variances reported for the estimates. This equality need not hold for the pseudo-counts inferred by the GL approach, which uses group counts in place of reported variances. 
This difference is discussed explicitly in the following sections.

\subsection{GL Newton Method}\label{sec:GL}
\begin{algorithm}[h!]
\caption{Greenland and Longnecker Algorithm}\label{alg:one}
\begin{algorithmic}[1] 
\REQUIRE $M_1, N, L$, Initialize $A$
\STATE $\mathrm{difference} \gets 1$
\WHILE{$\mathrm{difference} \geq 1e-4$}
  \STATE $A_{+} \gets \mathrm{sum}(A)$
  \STATE $a_0 \gets M_1 - A_{+}$
  \STATE $b_0 \gets n_0 - a_0$
  \STATE $B \gets N_{+} - A$
  \STATE $c_0 \gets \frac{1}{a_0} + \frac{1}{b_0}$
  \STATE $c \gets \frac{1}{A} + \frac{1}{B}$ 
  \hfill \COMMENT{Element-wise inverse}
  \STATE $e \gets L + \log(a_0) + \log(B) - \log(A) - \log(b_0)$ 
  \hfill \COMMENT{Element-wise $\log$}
  \STATE $H \gets$ matrix of size $n\times n$ whose diagonal elements are $c + c_0$ and whose off-diagonal elements are $c_0$
  \STATE $A \gets A + H^{-1}e$
  \STATE $\mathrm{difference} \gets \norm{H^{-1}e}_2$
\ENDWHILE
\end{algorithmic}
\end{algorithm}

The GL approach uses reported estimates, total  counts, and the total number of cases to find pseudo-counts in each category to match reported log-OR or log-RR estimates using an iterative root-finding method given in Algorithm~\ref{alg:one}. 
Indeed, Algorithm~\ref{alg:one} is exactly Newton's method for root-finding, applied to find pseudo-counts such that plug-in estimates from the pseudo-counts match those of the adjusted estimates reported in the original study. 
Once $A, B, a_0, b_0$ are found, the \cite{Greenland1992MethodsFT} uses these values to calculate the correlation coefficient $r_{ij}$ on log OR estimates $L_i$ and $L_j$ using~\eqref{eq:corrOR},
as well as covariances 
\[
    C_{ij} = r_{ij}\paren{V_i V_j}^{1/2}.
\]
 
For an arbitrary multi-variable function $f: \R^n \to \R^n$, the Newton iteration is given by
\begin{equation}\label{eqn:one}
    x_{k+1} = x_k - \braces{J_f(x_k)}^{-1}f(x_k)
\end{equation}
where $J_f$ is defined to be the Jacobian  matrix of $f$, comprising partial derivatives \citep{GautschiNumA}. Newton's method is {\it locally convergent}; meaning that when the initial iterate $x_0$ is ``close enough" to a root,~\eqref{eqn:one} will eventually find it; however, getting close enough can be tricky \citep{Süli_Mayers_2003}.  {\it Global convergence} refers to be ability of the algorithm to converge regardless of initialization.  \cite{Greenland1992MethodsFT} do not prove global convergence guarantees; and in fact as given in~\cite{Greenland1992MethodsFT} and summarized in Algorithm~\ref{alg:one}, the method can break depending on initialization. 

The function $g:\mathbb{R}^n \rightarrow \mathbb{R}^n$ whose zero we are searching for
appears in line 11 of Algorithm~\ref{alg:one}
and is given by
\begin{equation}\label{gradient}
g(A) = -L - \log(a_0 (A))\mathbf{1} - \log(B(A)) + \log(A) + \log(b_0 (A))\mathbf{1}.
\end{equation}
where $\mathbf{1}\in\mathbb{R}^n$ is the vector of ones of the right dimension, copying the values of the scalar quantity to all coordinates. 
By construction, $a_0, B$, and $b_0$ are all functions of $A$.
The Jacobian matrix $H$ is contains all the partial derivatives of $g(A)$ and is computed  in Algorithm ~\ref{alg:one}. \cite{Greenland1992MethodsFT} suggest using crude estimates to initialize $A$ if available, and otherwise using the null expected value: $M_1\frac{N_+}{\mathrm{sum}(N)}$. 
A priori, convergence is not guaranteed. 
In Section~\ref{sec:numerics}, we explore failure modes of existing implementations. 

We show in Section~\eqref{Cvx:GL} that the function $g$ in~\eqref{gradient} is the gradient of a convex function and recast the rootfinding problem for $g$ into a convex optimization problem which allows us to robustly compute the GL estimator. 
This leads to a variety of algorithms with global convergence guarantees, and more simply, to a DCP approach that does not need user-specified initialization and leverages the state-of-the-art open source optimization software \verb|cvxpy| \citep{diamond2016cvxpy}; we make this available to the community. 

\subsection{Hamling Method}\label{sec:hamling}
\cite{hamling2008facilitating} extended the work of~\cite{Greenland1992MethodsFT}, and also 
construct pseudo-counts $A, B, b_0,$ and $a_0$ using an iterative root-finding method. 
Once the pseudo-counts are obtained, the correlations across treatment effect exposures and overall covariance matrix are calculated exactly the same as by~\cite{Greenland1992MethodsFT}. 
The main difference is that~\cite{hamling2008facilitating} only requires estimates and their variances, along with $p$ and $z$ from Table~\ref{table:notation}, discussed in detail below. 

The two pieces of information that Hamling requires in addition to estimates and variances are $p$ and $z$, which
correspond to the ratio of unexposed controls to total number of controls, and the ratio of total number of controls to total number of cases, respectively.  
These quantities can be obtained by using crude reported estimates from the study, or from another pathway (e.g. literature) if the study did not report the quantities.  

\cite[Appendix A]{hamling2008facilitating} solve for $A, B, p', z'$ in terms of $a_0$ and $b_0$:
\begin{align}\label{def:hamling}
\begin{split}
    A_i &= \frac{1 + \frac{a_0 L_i}{b_0}}{V_i - \frac{1}{a_0} - \frac{1}{b_0}}, \quad
    B_i =  \frac{1 + \frac{b_0}{a_0 L_i}}{V_i - \frac{1}{a_0} - \frac{1}{b_0}}, \quad
    p' = \frac{b_0}{\sum_{i=1}^{n}B_i}, \quad
    z' = \frac{\sum_{i=1}^{n}B_i}{\sum_{i=1}^{n}A_i}.
\end{split}
\end{align}
The quantities $p'$ and $z'$ are functions of $(a_0, b_0)$, and the main idea of the Hamling method is to match 
$p'$, $z'$ to the $p, z$ values provided by the study, minimizing the squared differences:
\begin{equation}
\label{eq:ls_pz}
  \paren{\frac{p - p'}{p}}^2 + \paren{\frac{z - z'}{z}}^2 
\end{equation}
The iteration of Hamling, summarized in Algorithm~\ref{alg:ham}, update $a_0$ and $b_0$ through the equations~\eqref{def:hamling}. Once $(a_0, b_0)$ are found, equations~\eqref{def:hamling} yield all needed pseudo-counts. It is not obvious from~\cite{hamling2008facilitating}, but a consequence of our work here is equivalent to showing that~\eqref{eq:ls_pz} can always be brought to $0$, for all feasible inputs.  

\begin{algorithm}[h!]
\caption{Hamling Algorithm}\label{alg:ham}
\begin{algorithmic}[1] 
\REQUIRE $p, z, L, v$, Initialize $a_0, b_0$
\STATE error $\gets$ 1.0
\WHILE{$\mathrm{error} \geq 1e-4$}
    \STATE $A_i(a_0, b_0) \gets 
    \left(1 + \frac{a_0 L_i}{b_0}\right)/\left(V_i - \frac{1}{a_0} - \frac{1}{b_0}\right)$
    \STATE $B_i(a_0, b_0) \gets  
    \left(1 + \frac{b_0}{a_0 L_i}\right)/
    \left(V_i - \frac{1}{a_0} - \frac{1}{b_0}\right)$
    \STATE $p'(a_0, b_0) \gets 
    b_0/\left(\sum_{i=1}^{n} B_i(a_0, b_0) \right) $
    \STATE $z'(a_0, b_0) 
    \gets 
    \left(\sum_{i=1}^{n}B_i(a_0, b_0)\right)/
    \left(\sum_{i=1}^{n}A_i(a_0, b_0)\right)$
    \STATE $\mathrm{error} \gets \left(\frac{p - p'}{p}\right)^2 + \left(\frac{z - z'}{z}\right)^2$
    \STATE $a_0, b_0 \gets$ Update   \hfill \COMMENT{Black Box Optimization routine to shrink error, e.g. Excel or Stata}
\ENDWHILE
\end{algorithmic}
\end{algorithm}

~\cite{hamling2008facilitating} suggest using the Excel \verb+Solve+ function as a black-box optimizer. However, the solution method turns out to be less important than the choice of equations and their initialization.  In Section~\ref{sec:solve}, we show that a modified but equivalent system of nonlinear equations for OR always has a solution.
In contrast, the original formulation does not have any such guarantees, and~\cite{hamling2008facilitating}
discuss the need to use different starting points  to ensure converge in specific instances. In Section~\ref{sec:numerics}, we give specific, simple examples where the method as given in Algorithm~\eqref{alg:ham} fail to converge to a solution for $a_0$ and $b_0$ (returning negative counts $A$ or $B$), while the method of Section~\ref{sec:solve} succeeds. 
In the RR case, 
we show that it is in fact possible for the Hamling approach to catastrophically fail, which we discuss in detail in Section~\ref{sec:solve}.

\section{Convex Optimization Formulation of GL}\label{Cvx:GL}
In this section, we develop a robust GL approach  
by establishing that $g(A)$ used in the root-finding Newton method of Algorithm~\eqref{alg:one} is the gradient of a convex function. 
We show that the convex model of interest is a sum of entropic distance functions for both log ORs and log-RRs.  
We begin with log ORs. 

\subsection{GL: Odds Ratios}\label{sec:entropic}
Recall the function $g(A)$ that is the focus of the Newton's root finding method proposed by~\cite{Greenland1992MethodsFT}:
\[
    g(A) = -L - \log(a_0(A))\mathbf{1} - \log(B(A)) + \log(A) + \log(b_0(A))\mathbf{1}.
\]
We  can find the integral $G$ of $g$ and obtain an
objective that corresponds to this gradient:
\begin{align}\label{newt:obj}
\begin{split}
     G(A) = -L^\top A &+ \paren{a_0(A)\log(a_0(A))
    - a_0(A)}
    + \sum_{i=1}^{n}\paren{B_i(A)\log(B_i(A)) - B_i(A)}\\
    &+\sum_{i=1}^{n}\paren{A_i\log(A_i) - A_i}
    + \paren{b_0(A)\log(b_0(A)) - b_0(A)}.
\end{split}
\end{align}
From here, note that we may equivalently solve for the optimal $A$ by minimizing the integrated $G$. This is equivalent to finding roots of $g$ as GL does, since $\nabla G(A) = g(A)$ and a function is at an optimal value precisely when its gradient is zero.

Recall that a convex function $G$ satisfies \citep{boyd2004convex}
\begin{equation}
\label{eq:convexity}
G(\lambda A_1 + (1-\lambda) A_2) \leq \lambda G(A_1) + (1-\lambda) G(A_2) \quad \mbox{for all} \quad 0 < \lambda < 1, \quad \mbox{and} \quad  A_1, A_2 \in \mathbb{R}^n. 
\end{equation}
A closely related property called {\it strict convexity} requires strict inequality in~\eqref{eq:convexity} for $A_1 \neq A_2$. 
For a function with continuous derivative, as in our case, the convex property and first-order Taylor series expansion of $G$ yield the differential characterization of convexity 
\begin{equation}
\label{eq:convex_derivative}
G(A_2) \geq G(A_1) + (A_2-A_1)^\top \nabla G(A_1) \quad \mbox{for all} \quad A_1, A_2 \in \mathbb{R}^n. 
\end{equation}
The characterization~\eqref{eq:convex_derivative} means that if $\nabla G(A_1) = 0$, then necessarily 
\[
G(A_2) \geq G(A_1) \quad \mbox{for all} \quad A_2 \in\mathbb{R}^n,
\]
that is, $g(A_1) = 0$ guarantees $A_1$ must be the global minimizer of $G$. Moreover, a strictly convex function $G$ cannot have more than one global minimum; otherwise, given two such minima, we can use the strict version of~\eqref{eq:convexity} to get a point with a lower value for e.g. $\lambda = 1/2$.

Finally, for a function with second continuous derivative,  non-negative eigenvalues of the Hessian for any $A$ in the domain is a sufficient condition for convexity. As already discussed in Section~\ref{sec:GL}, the Jacobian matrix $H$ of $g$, which is exactly the Hessian of $G$, is symmetric positive definite, meaning all eigenvalues are actually positive, which means $G$ must be {\it strictly} convex \citep{boyd2004convex}. 

Putting these facts together, the root-finding problem for $g$~\eqref{gradient} is equivalent to minimizing  a strictly convex minimization problem with objective $G$~\eqref{newt:obj}. This perspective reveals that the original GL method can be strengthened by using additional structure and safeguards provided by $G$. For example, the simplest safeguard for Newton's method when minimizing $G$ is a step size search that moves in the Newton direction just enough to guarantee a proportional decrease $G$, and adding this element to Algorithm~\ref{alg:one} would already provide global convergence guarantees. The optimization problem is given by 
\begin{equation}\label{opt:main}
    \min_{ 0 \leq A \leq N} \quad G(A) 
\end{equation}
where $G(A)$ is given in~\eqref{newt:obj}. 
This formulation implicitly maintains domain constraints, that is, non-negativity of $A$, $N-A$, as well as non-negativity of $a_0$ and $b_0$, since the logarithm is only defined on $\mathbb{R}_+$. 
The key element in~\eqref{newt:obj} is the entropic distance function $f:[0,\infty)^m \to \R$:
\begin{equation}\label{entropic_fn}
    f(x) = x\log (x) - x .
\end{equation}
As we approach $0$, $x\log(x)$ goes to $0$, as can be easily seen by using L'Hôpital's rule.  
As $x$ grows large, $x\log(x)$ grows faster than $x$, so $f(x) \to \infty$ as $x \to \infty$. Finally the entropic function has positive second derivative on its domain
\[
f''(x) = 1/x
\]
so it is strictly convex. Since the sum of convex and strictly convex functions are strictly convex by definition, the entire objective $G$~\eqref{newt:obj} is strictly convex. 
This implies that any minimizer of $G$~\eqref{newt:obj} must be unique, and it remains to show only that a minimizer exists for $G$. 
\begin{theorem}\label{thm:exist}
    Suppose $N_+ > A, M_1 > 1^\top A,$ and $n_0 > a_0$ according to the variables defined in Table~\eqref{table:notation}. Let $L$ represent log ORs for the necessary exposure levels and take the elements of $L$ to be finite. Then the function $G(A)$~\eqref{newt:obj} always has a unique global minimizer.
\end{theorem}
For the proof, please see the Appendix~\ref{Appendix}. From Theorem~\ref{thm:exist}, $G$ always has a unique minimizer for feasible inputs, undergirding the approach of \cite{Greenland1992MethodsFT}.  
A unique global minimum exists under simple assumptions about problem data, and standard optimization solvers (including gradient, Gauss-Newton, quasi-Newton, and Newton), when properly safeguarded by trust region or line search,  will converge to the unique global minimum of $G$ for any feasible initialization of $A_0$. In particular, we use disciplined convex programming~\citep{grant2006disciplined}
to solve the problem. 
In Section~\ref{sec:numerics}, we show that the root-finding scheme of \cite{Greenland1992MethodsFT} is fragile with respect to initialization, but the new approach is guaranteed to work.

\subsection{GL: Relative Risk}
\label{sec:RR}
We now discuss the changes to apply the approach to log-RR scores. The overall approach and notation (see Table~\eqref{table:notation}) largely follow the development in the preceding section. $R$, the log RR score, 
is a function of problem data as given by~\cite{Greenland1992MethodsFT}:
\[
    \exp(R) = \frac{An_0}{N_+ a_0}, \quad 
    R = \log(A) - \log(N_+) - \log(a_0) + \log(n_0).
\]
Here, $N_+$ and $n_0$ are treated as known quantities, again following~\cite{Greenland1992MethodsFT}. To recover the pseudo-counts, we look for $A, a_0$ that are roots of
\begin{equation}\label{rr:newtgrad}
    h(A) = -R + \log(A) - \log(N_+) - \log(a_0 (A)) + \log(n_0).
\end{equation}
\cite{Greenland1992MethodsFT} suggest an algorithm similar to Algorithm~\eqref{alg:one} to construct cell counts for $A$ and $a_0$. 
Just as in Section~\ref{sec:entropic}, we cast this root-finding method as a way to solve a convex optimization program based on entropic distance, analogous to~\eqref{newt:obj}. 
Integrating Equation~\eqref{rr:newtgrad}, we obtain
\begin{equation}\label{obj:RR}
    H(A) = A^\top \paren{-L_R - \log(N_+) + \textbf{1}\log(n_0)}
    + \sum_{i=1}^n A_i \log(A_i) - A_i 
    + a_0(A)\log(a_0(A)) - a_0(A).
\end{equation}
The function $H$ is strictly convex, since it is the sum of three linear terms, and $n+1$ entropic distance functions (see the discussion in Section~\ref{sec:entropic}). 
We prove a theorem analogous to Theorem~\ref{thm:exist}, showing the existence of a solution under simple assumptions; uniqueness follows from strict convexity. 
\begin{theorem}\label{thm:exist2}
    Suppose $M_1 > 1^\top A$. Let $L_R$ represent log RR ratios for the necessary exposure levels such that $L_R$ is finite. Then the function $H(A)$ \eqref{obj:RR} always has a unique minimizer.
\end{theorem}
The proof for Theorem~\ref{thm:exist2} is in the appendix. In this way, we may construct the optimization problem
\begin{equation}\label{opt:rr}
    \min_{A\in \mathbb{R}_+^n} \quad H(A)
\end{equation}
where $H(A)$ is given in~\eqref{obj:RR}. 
By Theorem~\ref{thm:exist2}, problem~\eqref{opt:rr} must have a minimizer. A solution to the optimization problem \eqref{opt:rr} may be found by using any number of optimization methods, and in particular, we can also use disciplined convex programming~\citep{grant2006disciplined} to solve~\eqref{opt:rr}, just as in Section~\ref{sec:entropic}.

It may seem a natural fact that root finding here corresponds to a convex objective, but in our experience this  is an exception rather than the rule.  To be clear, while  minimizing a smooth convex function is often solved by a root-finding procedure on the gradient, the converse rarely holds, that is, a typical root finding problem rarely turns out to correspond to the gradient of a convex model. 
Case in point: when we consider the Hamling method, we do not have a convex interpretation, and as a result have to essentially use brute force to derive theoretical convergence guarantees. It is also quite fortunate that the convex reformulation works in a very similar way for the GL approach for both RR and OR. Again returning to Hamling, in the case of OR, we can find a counter-example guaranteed to fail. 
The contrast of GL with Hamling here underscores the rarity of the discovered relationship of the GL approach to convex minimization.

\section{Solvabililty of Hamling Method}\label{sec:solve}

In Section~\ref{sec:hamling}, we gave a brief overview of the method of \cite{hamling2008facilitating}, which involved formulating and solving nonlinear equations~\eqref{def:hamling} for $A_i$ and $B_i$. 
The approach relies on the reported variances rather than group totals  to infer pseudo-counts. Besides the estimates and variances, the Hamling approach needs only $p$ and $z$, see Table~\ref{table:notation}. However, the parametrization using  variances make the nonlinear equations of Hamling far more difficult to analyze than the GL approach. The original work~\cite{hamling2008facilitating} did not provide any guarantees, and in fact the authors' numerical examples suggest initialization may be quite  important. 
In this section, we prove that for the OR case, the equations always have a unique positive solution, and when properly initialized, the solution can always be found. In the RR case, the situation is more difficult; we present a counter-example where a solution to the Hamling equations cannot exist, and a partial theoretical result by deriving a sufficient condition for the existence of a solution to Hamling RR in the equivariant case. 

\subsection{Hamling: Odds Ratios \label{sec:hanlingOR}}

The quantities that~\cite{hamling2008facilitating} use, as functions of the underlying pseudo-counts, are given by: 
\begin{equation}
\label{eq:all_terms}
\begin{aligned}
R_i = \frac{A_ib_0}{a_0 B_i}, \quad
V_i  = \frac{1}{a_0} + \frac{1}{b_0} + \frac{1}{A_i} + \frac{1}{B_i}, \quad 
p = \frac{b_0}{\sum_{i=0}^n B_i}, \quad 
z  = \frac{\sum_{i=0}^n B_i}{\sum_{i=0}^n A_i}. 
\end{aligned}
\end{equation}
Using the substitution $B_i = \frac{A_i B_0}{A^0 R_i}$~\cite{hamling2008facilitating} 
obtains $B_i$ and $A_i$ in terms of $a_0$, $b_0$, $R_i$ and $V_i$:
\[
\begin{aligned}
B_i(a_0, b_0) &= \left(1+\frac{b_0}{a_0 R_i}\right) / \left(V_i - \frac{1}{a_0} - \frac{1}{b_0}\right) \\
A_i(a_0, b_0) &= \left(1+\frac{a_0R_i}{b_0}\right) / \left(V_i - \frac{1}{a_0} - \frac{1}{b_0}\right)
\end{aligned}
\]
Note that these equations for $A_i, B_i$ are the equations that~\cite{hamling2008facilitating} solve for, in terms of $a_0, b_0$, in order to match the variances of the pseudo-counts to the reported variances. Though, the authors do not solve these equations explicitly, instead using Algorithm~\ref{alg:ham} to estimate the changing parameter values $a_0,b_0$ and update pseudo-counts accordingly. 
\[
B_+ = \sum_{i=1}^n B_i, \quad A_+ = \sum_{i=1}^n A_i.
\]
Summing across each set of equations for $A_i$ and 
$B_i$ we get
\[
\begin{aligned}
    B_+ &= \sum_{i=1}^n \left(1+\frac{b_0}{a_0R_i}\right) / \left(V_i - \frac{1}{a_0} - \frac{1}{b_0}\right) \\
    A_+ &= \sum_{i=1}^n \left(1+\frac{a_0R_i}{b_0}\right) / \left(V_i - \frac{1}{a_0} - \frac{1}{b_0}\right) \\
\end{aligned}
\]
From the definitions of $p$ and $z$ we  have 
\[
\begin{aligned}
B_+  = \frac{1-p}{p}b_0, \quad
A_+  = \frac{1}{z(1-p)}B_+ - a_0 = \frac{1}{zp} b_0 - a_0.
\end{aligned}
\]
Combining these equations together, we get a system of two explicit equations for unknowns $a_0$ and $b_0$: 
\begin{equation}
\label{eq:nlEqs}
\begin{aligned}
    \frac{1-p}{p}b_0 &= \sum_{i=1}^n \left(1+\frac{b_0}{a_0 R_i}\right) / \left(V_i - \frac{1}{a_0} - \frac{1}{b_0}\right) \\
    \frac{1}{zp} b_0 - a_0 &= \sum_{i=1}^n \left(1+\frac{a_0R_i}{b_0}\right) / \left(V_i - \frac{1}{a_0} - \frac{1}{b_0}\right) \\
\end{aligned}
\end{equation}


The approach developed here is  similar in nature to that of~\cite{hamling2008facilitating}, but 
equations~\eqref{eq:nlEqs} are not derived in~\cite{hamling2008facilitating}.  The explicit form of~\eqref{eq:nlEqs} is used to prove the results below, namely that equations~\eqref{eq:nlEqs} always have a solution. 

First, we show that a unique positive solution to~\eqref{eq:nlEqs} exists when all the variances are identical, that is, all $V_i = v$. The theorem for this case serves as a base case for the induction in the general result, and also is of interest since the proof technique is direct; we actually find the closed form of the solution. 

\begin{theorem}
\label{thm:equivariant}
Suppose all of the $V_i$ are equal to the scalar $v >0$. 
Then there is a unique positive solution of the equations~\eqref{eq:nlEqs} for any value $p \in (0,1)$ and any value of $z > 0$, and any set of positive 
estimates $R_i$. 

Let $c = \frac{a_0}{b_0}$. Let $r_1 = \sum_{i=1}^n \frac{1}{R_i}$
and $r_2 = \sum_{i=1}^n R_i$. 
Then the positive solution to Hamling is given by 
\[
c = \frac{n p z - n z + n - p r_1 z + \sqrt{D}}{2 z (n p +(1- p) r_2)}
\]
where  
\[
D = n^2 p^2 z^2 - 2 n^2 p z^2 + 2 n^2 p z + n^2 z^2 - 2 n^2 z + n^2 - 2 n p^2 r_1 z^2 + 2 n p r_1 z^2 + 2 n p r_1 z + p^2 r_1^2 z^2 - 4 p r_1 r_2 z + 4 r_1 r_2 z.
\]
Once we have $c$, the solutions to~\eqref{eq:nlEqs} are given by 
\[
b_0   = \frac{1}{v}\left(\frac{p}{1-p}\left(n + \frac{r_1}{c}\right) + 1 + \frac{1}{c}\right), \quad 
a_0 = cb_0. 
\]

\end{theorem} 
We provide a proof in Section~\ref{pf:equivariant} in the Appendix. The crux of the proof is to show that $D$ is always positive, for any feasible inputs $(n, p, z, r_1, r_2)$. 
An interesting consequence of the proof is that in addition to the unique positive solution for $c$ (and hence $a_0)$, there is also a unique negative solution for $c$ and $a_0$, obtained by taking the negative branch in the quadratic formula. From our numerical experience with Hamling, both our implementation and the one in \verb{dosresmeta{ can find the negative $a_0$ solution, leading to infeasible pseudo-counts, when incorrectly initialized.

We now show by induction that equations~\eqref{eq:nlEqs} always have a unique feasible solution in the general case.

\begin{theorem}\label{thm:solution}
For any set of positive $V_i$, positive $R_i$, $p \in (0,1)$ and $z> 0$,  the equations~\eqref{eq:nlEqs}  have a positive solution with $a_0 >0$ and
$b_0 > 0$. 
\end{theorem} 

See the Appendix, Section~\ref{pf:solution} for a proof of Theorem~\eqref{thm:solution}. The proof proceeds by induction, as it is impossible to find a closed form solution in the general case. This result ensures convergence to a tuple $(a_0,b_0)$ that can be used to construct the cell counts $A$ and $B$ according to equations~\eqref{def:hamling}. Our presentation of the nonlinear system in the form of equations~\eqref{eq:nlEqs} provides robustness to the method of \cite{hamling2008facilitating}, guaranteeing solutions 
for any choice of positive $V_i$. The inductive step of the theorem shown in Section~\eqref{pf:solution} of the Appendix uses the structure of the equations to show the existence of the solution.

To find the solution in practice, we minimize the squared norm of the equations~\eqref{eq:nlEqs}, similar to the approach shown in Algorithm~\ref{alg:ham}. The construction of the proof assumes the positivity of the denominators $V_i - \frac{1}{a_0} - \frac{1}{b_0}$ throughout. This guides our initialization strategy to ensure that $a_0$ and $b_0$ are large enough that positivity holds for the smallest reported variance, 
\[
V_{\min} = \min_{i} V_i.
\]
From a theoretical standpoint, the nonlinear constraint 
\[
a_0 + b_0 \leq \epsilon a_0 b_0 V_{\min}
\]
may be needed and can be maintained via line search, in practice the method always converges as long as the constraint holds at initialization. This is a markedly different strategy than the one suggested by~\cite{hamling2008facilitating}, who focus on $p'$ and $z'$ computed from total counts in the data. 
Their strategy, as implemented by~\cite{dosresmeta}, 
fails for cases where the reported variances are small, and is discussed in Section~\ref{sec:numerics}.

\subsection{Hamling: Relative Risk\label{sec:hamlingRR}}
We now consider the log RR scores. We use the exact same notation as what has been described in the current section, except we now use $L_R$ to imply the log RRs instead of log ORs. We have
\[
R_i = \frac{A_i b_0}{a_0 B_i}
\]
where, following the notation of~\cite{hamling2008facilitating}, $b_0$ now indicates total subjects in the reference group, and $B_i$ indicates total subjects in each risk group, with $a_0$ reference non-cases and $A_i$ cases across exposure groups. Thus we have $b_0 > a_0$ and $B_i > A_i$. Moreover, from classic results we have 
\[
V_i = \frac{1}{a_0}  - \frac{1}{b_0} + \frac{1}{A_i} - \frac{1}{B_i}.
\]
This becomes the key difference that underlies the construction of our new equations. Indeed, we obtain
\begin{align}\label{eq:hamR}
\begin{split}
A_i &= \frac{1 - \frac{a_0 {R_i}}{b_0}}{V_i - \frac{1}{a_0} + \frac{1}{b_0}}\\ 
B_i &= \frac{\frac{b_0}{a_0 {R_i}} - 1}{V_i - \frac{1}{a_0} + \frac{1}{b_0}}.
\end{split}
\end{align}
The constraint that $B_i > A_i$ doesn't give any new information, since it is equivalent to 
\[
\frac{b_0}{a_0{R_i}} + \frac{a_0 {R_i}}{b_0}> 2.
\]
The sum of any positive quantity and reciprocal is always greater  than or equal to $2$, with the minimum attained when the quantity is exactly $1$. We do however know something about $z$. Recall the formulas
\[
p = \frac{b_0}{\sum_{i=0}^n B_i}, \quad
z  = \frac{\sum_{i=0}^n B_i}{\sum_{i=0}^n A_i}. 
\]
For the relative risk case, by definition we have $z \geq 1$. 

Using  equations~\eqref{eq:hamR}  and formulas for $p$ and $z$, we construct the two nonlinear equations that are analogous to equations~\eqref{eq:nlEqs}:
\begin{equation}
\label{eq:nlEqsLogs}
\begin{aligned}
    \frac{1-p}{p}b_0 &= \sum_{i=1}^n \left(\frac{b_0}{a_0 R_i} - 1\right) / \left(V_i - \frac{1}{a_0} + \frac{1}{b_0}\right) \\
    \frac{1}{zp} b_0 - a_0 &= \sum_{i=1}^n \left(1-\frac{a_0R_i}{b_0}\right) / \left(V_i - \frac{1}{a_0} + \frac{1}{b_0}\right) .
\end{aligned}
\end{equation}
We now give analogous results to Theorems~\eqref{thm:equivariant} and \eqref{thm:solution}, which are proved in the Appendix. 

\begin{theorem}
\label{thm:equivariantRR}
Suppose all of the $V_i$ are equal to the scalar $v >0$. 
Then there is a unique positive solution of the equations~\eqref{eq:nlEqsLogs} for any value $p \in (0,1)$ and any value of $z > 0$, and any set of positive estimates $R_i$, if and only if
\[
(1-p)z \geq \left(\frac{1-p}{p}\right)^24r_2, \quad (1-p)z \geq 1. 
\]
Using the notation of Theorem~\ref{thm:equivariant}, when the conditions above are satisfied the positive solution $c = \frac{a_0}{b_0}$ is given by 
\[
c = \frac{n(z-pz + 1) + p r_1 z + \sqrt{D}}{2 z (n p +(1-p) r_2)}
\]
where 
\[
D = (n(pz-z-1)-r_1zp)^2-4r_1(nzp+r_2z-r_2pz). 
\]
Once $c$ is found, we have 
\[
b_0 = \frac{1}{Vc}\left(\frac{p}{1-p} (r_1 - cn) + (1-c)\right), \quad a_0 = cb_0. 
\]

\end{theorem} 
A simple counter-example that violates the two inequalities required by Therorem~\ref{thm:equivariantRR}, and for which there is no solution, is given by 
\[
R_1 = 0.9328, R_2 = 0.062, p = 0.1, z = 1.1. 
\]
These values have no solution in the RR example for any equal variance values  $V_1 = V_2 = v$.  In Section~\ref{sec:numerics}, we show that available implementations return nonsensical results, and in fact cannot solve the defining equations, which makes sense, given that $D <0$ in this case. In contrast to the previous section, there is no way to fix this issue, a solution simply cannot exist. The best we can do in such a case is to suggest the modeler check their inputs $R_i, V_i, p, z$ or consider using the GL approach, which is always guaranteed to work.


\section{Numerical Examples}\label{sec:numerics}
In this section, we review detailed examples of the implementation and results of our proposed methods as described in Sections~\ref{Cvx:GL} and~\ref{sec:solve}. First, we show that the corrected methods we proposed reproduce the results of \cite{Greenland1992MethodsFT} and \cite{hamling2008facilitating} for the canonical examples in these papers. Second, we show failure modes for \cite{Greenland1992MethodsFT} and \cite{hamling2008facilitating} and correct estimates 
from the robust implementations using the results in this work. 
For GL, we leverage the connection to convex optimization to provide software using disciplined convex programming libraries \verb|CVXPY| for our modification of the approach of \cite{Greenland1992MethodsFT}. For Hamling, we use \verb|SciPy| optimization routines with a theoretically justified initialization to solve the root finding problem. 
To demonstrate the failure modes, we use the  \verb|R| library \verb|dosresmeta| \citep{dosresmeta}, which implements both GL and Hamling methods. 


\subsection{Results Comparison to GL and Hamling: Canonical Examples}
We use the data from  \cite[Table 1]{Greenland1992MethodsFT} as a simple example showing that the optimized GL reproduces the same results as  regular~\cite{Greenland1992MethodsFT} and \cite{hamling2008facilitating} for simple problems. In the example given by \cite{Greenland1992MethodsFT}, the authors fit the linear-logistic model
\[
    \lambda(x,z) = \alpha  + \beta x.
\]
In this case, the model is giving the log-odds of a subject being a case, and we want to estimate $\beta$. The data $x$ represent alcohol intake as exposure levels. We present a summary of the adjusted estimates we obtain using our convex formulation for the objective of \cite{Greenland1992MethodsFT} and solutions to the modified system of equations originally in \cite{hamling2008facilitating}, and showing the coefficient value $\Hat{\beta}$ estimate along with the variance estimate for each method.

In Table~\ref{table:odds-est} we present the least-squares estimates generated from the four different types of pseudo-count fitting techniques described in this study. Denote by ``Unadjusted" as using reported variances with the independence assumption. Denote by ``GL" the least-squares and variance estimates obtained by the cell-fitting procedure of \cite{Greenland1992MethodsFT}. Denote by ``Hamling" the estimates produced from the method of \cite{hamling2008facilitating}. Denote by ``Convex GL" as the estimates obtained from our fitting procedure that modifies the method of \cite{Greenland1992MethodsFT} as described in Section~\ref{Cvx:GL}. Denote by ``Solved Hamling" as the estimates obtained from our fitting procedure that modifies the method of \cite{hamling2008facilitating} as described in Section~\ref{sec:solve}.

\begin{table}[h!]
\caption{Estimates and variances table--log-odds ratios\label{table:odds-est}. }
\begin{center}
\begin{tabular}{|c|c|c|} \hline
Method & $\Hat{\beta}$ & Variance  \\ \hline 
Unadjusted & 0.0334 & 0.000349 \\ 
\hline
GL & 0.0454 & 0.000427 \\
\hline
Convex GL & 0.0454 & 0.000427 \\
\hline
Hamling & 0.04588 & 0.000421 \\
\hline
Solved Hamling & 0.04588 & 0.000421 \\
\hline
\end{tabular}
\end{center}
\end{table}

The Convex GL method produces the same results as the original GL approach~\cite{Greenland1992MethodsFT} when the latter succeeds. Additionally, our Solved Hamling method produces the same results as the standard Hamling method when the latter succeeds. There are numerical differences in variance results for corresponding methods; for the Convex GL approach that uses DCP, we use a high degree of precision in the solver, so these results correspond to solving the equations to a greater degree of precision. 
The estimates obtained by Hamling vs. GL differ, but 
this is to be expected, as discussed in Section~\ref{sec:cor_cov}.

\begin{figure}[htbp]
  \centering
  \includegraphics[width=0.8\textwidth]{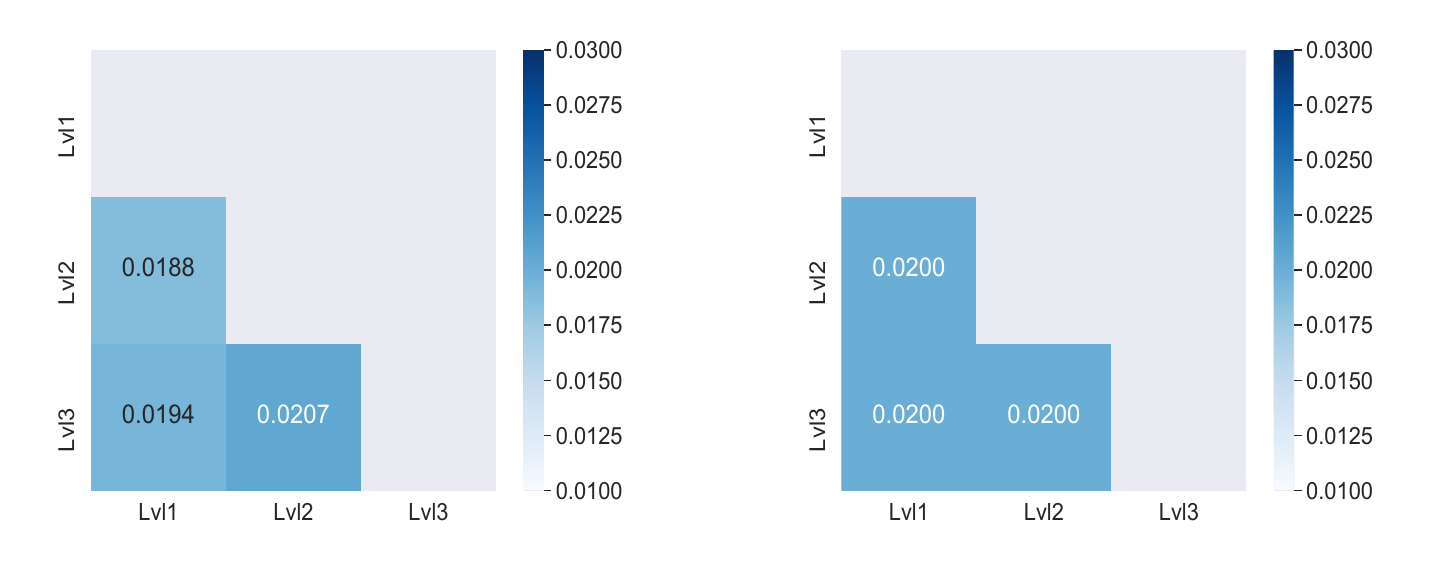} 
  \caption{Estimated covariance matrices for OR. Left: Covariance matrix generated from the Convex GL pseudo-counts; Right: Covariance matrix generated from the Solved Hamling pseudo-counts. }
  \label{fig:cov_matrix_OR}
\end{figure}

We next include a summary of the pseudo-counts only of cases generated by each method in Table~\ref{table:pseudoOR}. We follow the same notation used in Table~\ref{table:notation} for cases.
\begin{table}[h!]
\caption{Pseudo-count table--log-odds ratios\label{table:pseudoOR}. }
\begin{center}
\begin{tabular}{|c|c|c|c|c|} \hline
Method & $a_0$ & $A_1$ & $A_2$ & $A_3$  \\ \hline 
GL & 160.4702 & 70.2046 & 95.4696 & 124.8556 \\
\hline
Convex GL & 160.5064 & 70.3304 & 95.4857 & 124.6776 \\
\hline
Hamling & 96.2699 & 50.9684 & 57.2220 & 67.7043 \\
\hline
Solved Hamling & 96.2653 & 50.9654 & 57.2180 & 67.6989 \\
\hline
\end{tabular}
\end{center}
\end{table}
On this simple example, the pseudo-counts for cases generated by our methods match closely those generated by the original methods. Again counts generated by GL methods differ from those generated by Hamling methods, since the GL relies on group counts, whereas Hamling uses reported variances. 
The pseudo-counts are an intermediate result whose main purpose is to obtain the covariance matrix, and we compare the covariance matrices obtained by these methods in  Figure~\ref{fig:cov_matrix_OR}.

There are small differences between the individual entries 
in Figure~\ref{fig:cov_matrix_OR}. 
These differences are in fact what cause the estimates in Table~\ref{table:odds-est} to vary slightly between the GL and Hamling-based methods. 
Note that the Convex GL covariance matrix has different entries, whereas the entries of the (Solved) Hamling covariance matrix are identical. This is due to the variance model in the construction of the Hamling estimators. 

Next, we run a similar test on RRs, using the
alcohol and colorectal cancer data and results in 
\cite{orsini2012meta}. We  present a summary of the adjusted estimates obtained by our methods and by the methods of GL and Hamling. We use data directly from \verb|dosresmeta|, specifically the \verb|alcohol_crc| dataframe, and analyze the subset id author \verb|atm|. In Table~\eqref{table:rel-risk} we present the least-squares estimates, similar to what was shown above.

\begin{table}[h!]
\caption{Estimates and variances table--log-relative risks\label{table:rel-risk}. }
\begin{center}
\begin{tabular}{|c|c|c|} \hline
Method & $\Hat{\beta}$ & Variance  \\ \hline 
Unadjusted & -0.00294 & 1.5865e-05 \\ 
\hline
GL & 0.0071 & 1.5176e-05 \\
\hline
Convex GL & 0.0071 & 1.5166e-05 \\
\hline
Hamling & 0.0063 & 1.5490e-05 \\
\hline
Solved Hamling & 0.0063 & 1.5436e-05 \\
\hline
\end{tabular}
\end{center}
\end{table}

Once again we see small numerical differences in variance estimates, with our estimates using high precision on the equation solves. We also see a larger difference between the estimates obtained by GL vs.  Hamling, a direct consequence of the different parametrizations. 

We provide a summary of case pseudo-counts generated by each method in Table~\ref{table:pseudoRR}. The pseudo-count estimates 
within method families are close; while counts between GL and Hamling methods match in some groups but differ in others, causing the differences observed in estimate values in Table~\ref{table:rel-risk}. 
\begin{table}[h!]
\caption{Pseudo-count table--log-relative risk\label{table:pseudoRR}. }
\begin{center}
\begin{tabular}{|c|c|c|c|c|c|c|} \hline
Method & $a_0$ & $A_1$ & $A_2$ & $A_3$ & $A_4$ & $A_5$ \\ \hline 
GL & 26.5957 &  34.0061 & 42.8532 & 33.3584 & 17.9492 & 29.2359 \\
\hline
Convex GL & 26.5973 & 34.0061 & 42.8532 & 33.3583 & 17.9492 & 29.2359 \\
\hline
Hamling & 26.4495 & 39.5129 & 44.2940 & 31.6140 & 15.3332 & 22.6277 \\
\hline
Solved Hamling & 26.4087 & 39.4526 & 44.2234 & 31.5706 & 15.3105 & 22.5738 \\
\hline
\end{tabular}
\end{center}
\end{table}

\begin{figure}[h!]
  \centering
  \includegraphics[width=0.8\textwidth]{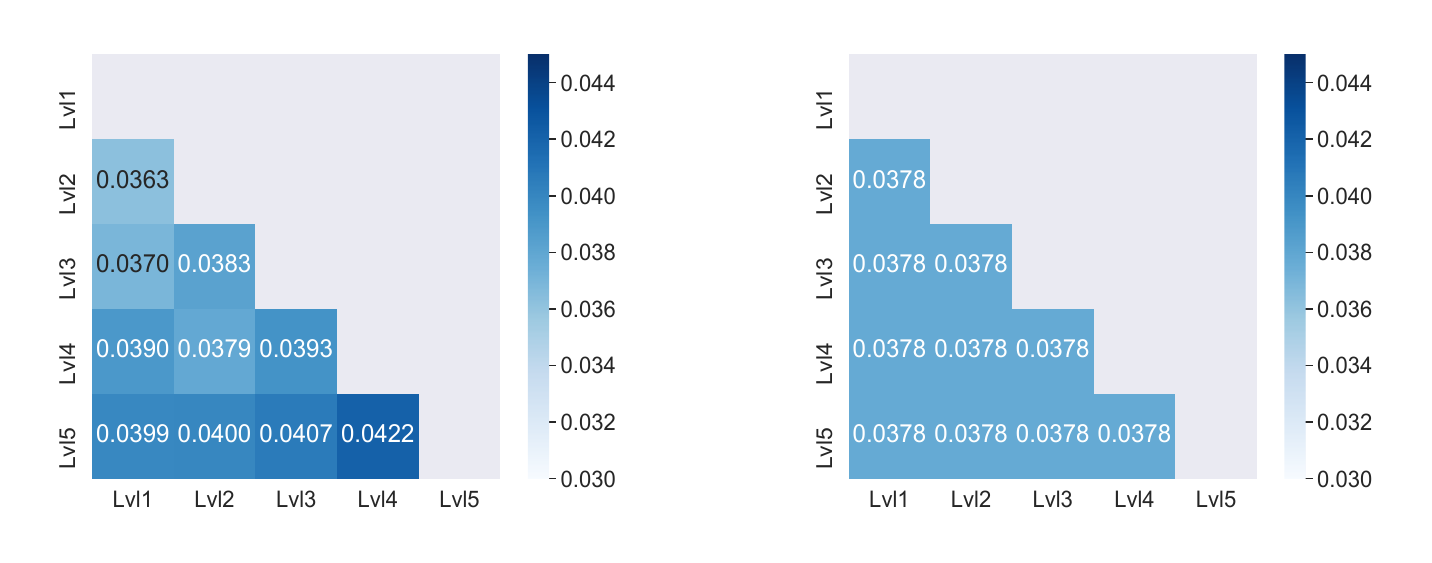} 
  \caption{Estimated covariance matrices for RR. Left: Covariance matrix generated from the Convex GL pseudo-counts; Right: Covariance matrix generated from the Solved Hamling pseudo-counts. Both cases are with respect to the relative risk regime.}
  \label{fig:cov_matrix_RR}
\end{figure}

Covariance matrices obtained from pseudo-counts generated by our  Convex GL and Solved Hamling methods are shown in Figure~\ref{fig:cov_matrix_RR}. We again see identical entries in the covariance matrix produced from Hamling. The differences in covariance values between the matrices explain the differences in estimates values in Table~\ref{table:rel-risk}.

We now continue to the avoidable failure modes, providing simple OR examples where the original GL and Hamling methods fail but our Convex GL and Solved Hamling methods succeed. 

\subsection{Original method failure and Corrected success}
In this section we produce simple failure modes for original GL and Hamling methods, and show that new methods work on these cases, as expected from the theoretical results. This is reassuring to practitioners running many analyses; the need to re-initialize current methods and potential quiet failures of the Hamling method can both be avoided with straightforward modifications. 
To demonstrate the failure modes, we perturb the \verb|alcohol_cvd| data from \verb|dosresmeta|.

\subsubsection{GL method failure}
\label{sec:GLfail}
Using the method of GL first, we change the number of subjects at each exposure level in the \verb|alcohol_cvd| dataset to be a function of the number of cases in the same dataset at the corresponding exposure levels. Namely, we  modify the number of subjects to be 
\[
    N = A + t
\]
for integer values $t = \{1,\dots,20\}$. 
The lower the $t$, the more extreme the situation, corresponding to very few controls in each group. 
We use each $N$ as input data to the standard GL routine to construct pseudo-counts using the GL method. For $t \leq 13$, the original GL method in \verb+dosresmeta+ fails. 

In the cases of failure, even though the initial $A$ is feasible, GL iterations run afoul of the 
logarithmic terms in the \verb+dosresmeta+ implementation for low $t$. For $t \geq 14$, this issue disappears. 
The entire problem is avoided when we use the convex GL approach, which succeeds in all cases. 



\begin{figure}[htbp]
  \centering
  \includegraphics[width=0.8\textwidth]{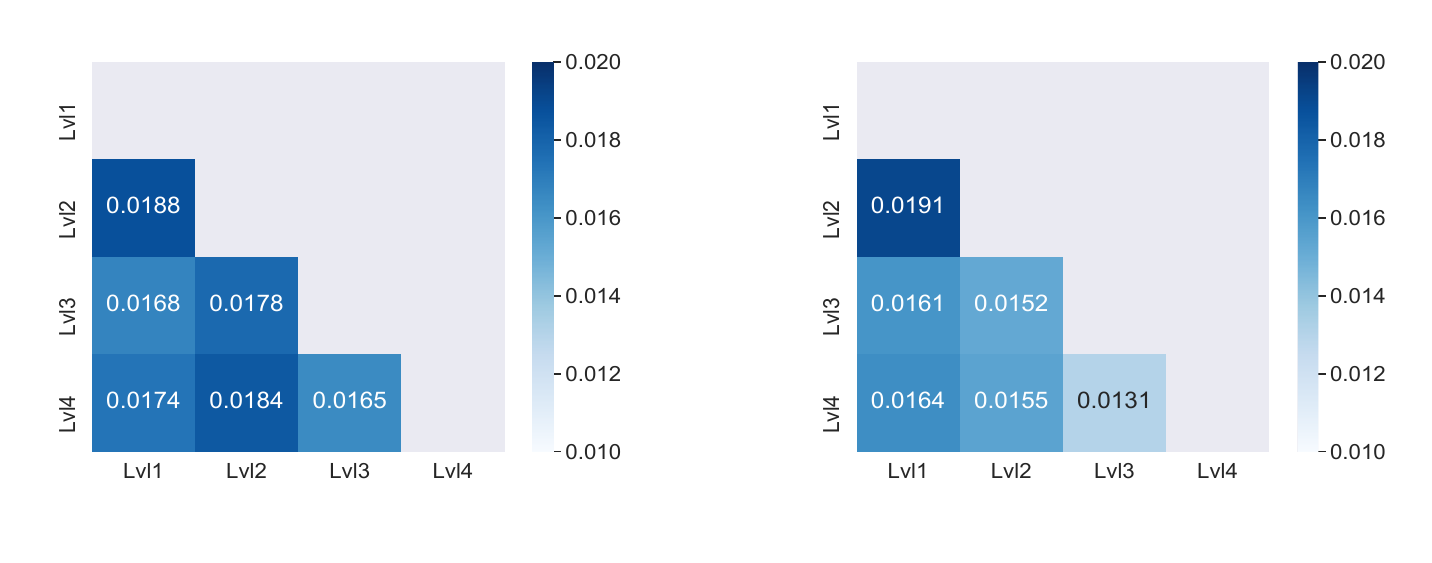} 
  \caption{Comparison of GL covariance matrices for original data vs. perturbed data. Left: Covariance matrix generated from the Convex GL pseudo-counts on original data for alcohol; Right: Covariance matrix generated from the Convex GL pseudo-counts on $N^1 = A+1$, a hypothetical where there is only one control in every group. The original GL method~\cite{Greenland1992MethodsFT} fails on the hypothetical example shown on the right.}
  \label{fig:cov_matrix_breakGL}
\end{figure}

The new Convex GL method succeeds even in the extreme case 
when $t=1$. We compare the covariance matrix for $N^1 = A + 1$ compared to the covariance matrix GL obtains on the original data in Figure~\ref{fig:cov_matrix_breakGL}. We see that the covariance matrices returned for the original and perturbed data are fairly close, suggesting that correlations are well-behaved in such cases and underscoring the need for a robust method. 
In other words, results for GL will likely be useful even for small studies when we have very few controls.  

\subsubsection{Hamling method failure}
The Hamling method fails when default initialization fails to guarantee positivity of all denominators $V_i - \frac{1}{a_0} -\frac{1}{b_0}$. For example, the Hamling initialization used by \verb{dosresmeta{ can break when the input $V_i$ are very small. In this case, the \verb+dosresmeta+ Hamling approach returns negative pseudo-counts, and correlations computed using these counts. 

To show this failure mode, we alter the \verb+alcohol_cvd+ dataset in \verb|dosresmeta| by changing the reported variances to be
\[
   \hat{V} =  \paren{\mathrm{NA}, 0.001,0.01,0.2,0.9}
\]
where the NA is a placeholder for the reference exposure level. Passing this data into the \verb|hamling| method in \verb|dosresmeta|, we obtain negative values in the estimated counts for cases and non-cases at the first level of exposure, as shown in Table~\ref{table:HamlingBreak}. 
\begin{table}[h!]
\caption{Pseudo-count table--broken Hamling Example \label{table:HamlingBreak}. }
\begin{center}
\begin{tabular}{|c|c|c|c|c|c|} \hline
Method & $a_0$ & $A_1$ & $A_2$ & $A_3$ & $A_4$ \\ \hline 
Hamling & 189.7 & {\bf -207.5} & 514.2 & 8.6 & 2.2  \\
\hline
Solved Hamling & 2897.8 & 2976.1 & 157.2 & 31.5706 & 9.2  \\
\hline
\end{tabular}
\end{center}
\end{table}

\begin{figure}[htbp]
  \centering
  \includegraphics[width=0.8\textwidth]{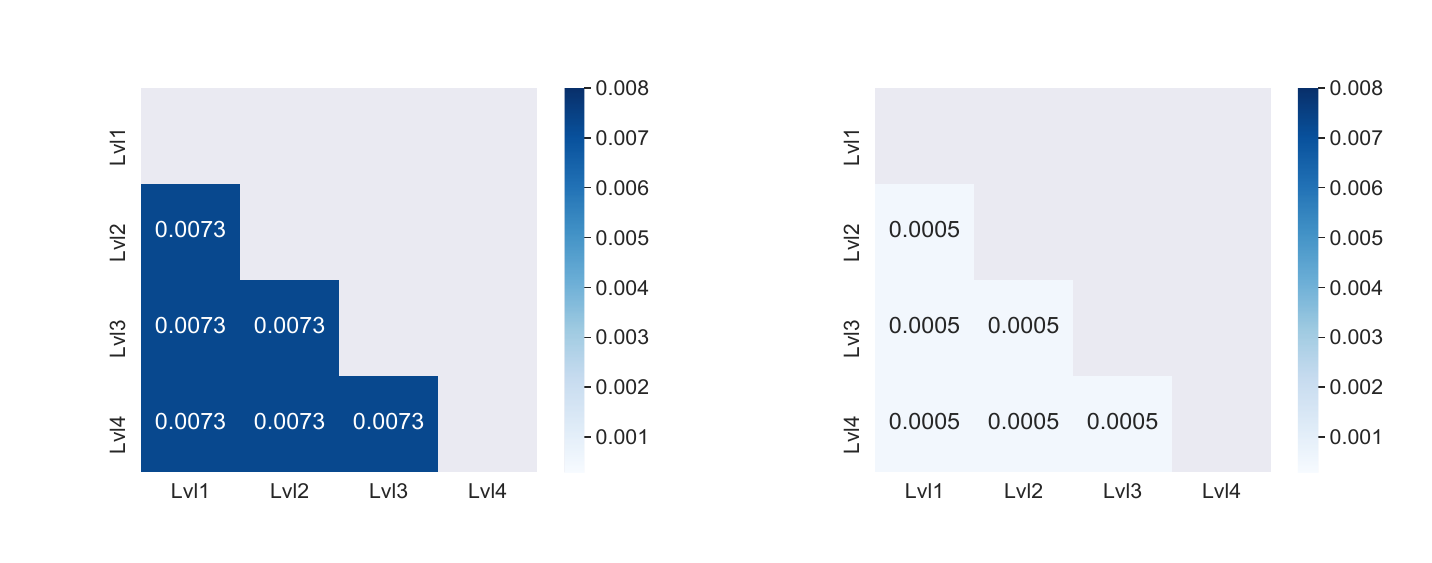} 
  \caption{Comparison between covariance matrices generated by Hamling from wrong pseudo-counts and  correct pseudo-counts. Left: Covariance matrix generated from the negative Hamling pseudo-counts; Right: Covariance matrix generated from the Solved Hamling pseudo-counts. The correct values result in a much smaller between-level covariance than the incorrect values in this example.}
  \label{fig:cov_matrix_break}
\end{figure}

The method of Hamling fails silently, since it then uses the negative values to compute the covariance matrix. To study the downstream effects,  we compare the covariance matrices constructed by \verb{dosresmeta{ from the wrong pseudo-counts generated by the original Hamling method with those generated by the solved Hamling method in Figure~\ref{fig:cov_matrix_break}. The solved Hamling method obtains an order of magnitude smaller correlation across the subgroups. This means that when Hamling fails quietly, it will provide estimates that deviate further from the uncorrected estimates compared to the correctly solved formulation.

We extend this example to study the range of the failure mode as a function of the scale of variance values. For simplicity we vary only the first element of $\hat{V}$. 
We then assess whether there are any negative values in the constructed pseudo-counts for cases $A$ by the Hamling method. As can easily be verified, the variance values below $10^{-3}$ in the first numerical coordinate produce negative values in $A$. Obviously for smaller estimates the method still fails, but such small variances correspond to huge sample sizes that are unlikely to occur in practice.  
The variance values greater than $10^{-3}$ produce only positive values, even beyond 1. This shows clearly how the Hamling method fails for small enough variance values given default initialization in \verb|dosresmeta|, and can be fixed easily by using the strategy discussed in Section~\ref{sec:solve}.


To fix the problem we use the initialization suggested by the theoretical analysis. 
Specifically, we construct the initialization parameters $a_0, b_0$ as
\[
    (a_0,b_0) = \paren{\frac{10}{\min(v)},\frac{10}{\min(v)}}.
\]
The underlying idea is that the large initialization ensures the denominators of $A_i$ and $B_i$ in equations~\eqref{def:hamling} remain positive, ensuring all counts are positive. This works well numerically, and does not break regardless of the $v_i$ values. 
This provides empirical support for the proof technique given in Theorem~\ref{thm:solution}. 

In the next section, we study the unavoidable failure mode of Hamling for RRs.

\subsection{Hamling Failure for RR\label{sec:ORfail}}

We review the counter-example presented in Section~\ref{sec:solve} 
\[
R_1 = 0.9328, R_2 = 0.062, p = 0.1, z = 1.1.  
\]
This example was obtained by violating the conditions presented in Theorem~\ref{thm:equivariantRR} for the equivariant case. The failure corresponds to obtaining a negative discriminant in the quadratic formula for the ratio $c = \frac{a_0}{b_0}$, and means that a solution cannot exist, regardless of reported (equal) variances. To see this bear out in practice we make a simple choice  
\[
v_1 = v_2 = 1.0.
\]
Running \verb{dosresmeta{ on this example gives us results in Table~\ref{table:RRfail}. 
\begin{table}[h!]
\caption{Hamling Results  for RR Counter-Example\label{table:RRfail}. }
\begin{center}
\begin{tabular}{|c|c|} \hline
 $A$ & $N$  \\ \hline 
 1.4 & 1.3 \\ 
\hline
$-1.1\times 10^{-5}$ & $-1.1\times 10^{-5}$ \\
\hline
$0.88$ & $13.3$ \\
\hline
\end{tabular}
\end{center}
\end{table}
We see negative values for $A$ and $N$, a problem for any situation, and $a_0 > b_0$, which is impossible for RR. These issues still can still occur for a candidate solution to the equations~\eqref{eq:nlEqsLogs}. However,
the claim we made is stronger, that is, a solution that satisfies 
the six equations corresponding to $p, z, R_1, R_2, v_1, v_2$ cannot exist. 
When we review the \verb{dosresmeta{ result with respect to these six equations, we find that in fact, two of the six are not satisfied: 
\[
\begin{aligned}
R_1(A,N) & = 0.93, \quad R_2(A,N) = 0.062,
 \quad v_2(A,N) = 1.0,  \quad p(A,N) = 0.1;  \\
 v_1(A,N) & =  -0.05; \quad z(A,N) =  6.33. 
\end{aligned}
\]
In contrast to the previous examples, there is no way to fix this; we know from the proof of Theorem~\ref{thm:equivariantRR} that no solutions can exist to this example. 

\section{Conclusion\label{sec:conclusion}}

In this paper we have taken a closer look at the 
methods of~\cite{Greenland1992MethodsFT} and~\cite{hamling2008facilitating}. 

We have shown that the GL approach lends itself to a reformulation to minimizing a convex model, for both ORs and RRs. In both cases we can avoid all numerical difficulty and guarantee convergence to the unique optimal point for any feasible data inputs. This was a rather surprising finding that initially motivated us to write the paper. 
The convex loss that emerged when we integrated the optimality conditions is the entropic distance function, an object that appears in other areas of mathematics and statistics. An unexplored consequence of the 
connection to convex models is that it is now easy to include side information (if such information is available to modelers) through the use of linear equality and inequality constraints on the pseudo-counts $A$. As long as there is a feasible $A$, the proof theory in this paper guarantees a unique solution, and modifying the formulation is straightforward in \verb{cvxpy{. We leave further exploration of this idea to future work.

For the Hamling method, the story is more complicated. In the case of OR, we were able to show that the Hamling equations always have a solution. In fact we obtained a closed form solution for the equivariant case (all reported variances equal) and provided a proof by induction for the general case. This means that literally for any observed ORs, variances, $p$, and $z$, we can always find a solution.

In contrast, for RR, there is no guarantee that Hamling will work. We presented a counter-example when there are only two alternative groups. Counter-examples are by nature odd, but  nonetheless there is a fundamental difference between RR and OR for Hamling stemming from relying on reported variances. This is curious. Between the methods of GL and Hamling, when faced with many meta-analyses we find the Hamling approach more appealing, since it only needs $p$ and $z$ in addition to reported estimates and variances. Based on the RR failure, we should keep the GL method available should an unavoidable failure mode arise.

We have done our best to make the results as interpretable and clear as possible. We have an implementation for GL and Hamling methods publicly available\footnote{https://github.com/ihmeuw-msca/CorrelationCorrection}; and we have shown simple cases where we can break the widely used \verb{dosresmeta{ package using simple examples. Using the insights in this paper,  safeguarding estimates available in other packages is a straightforward task. 
For GL, it is a matter of providing standard optimization guardrails, such as a line search. For Hamling, it is a change in the initialization strategy based on the minimum reported variance.

\pagebreak
\bibliographystyle{plainnat}
\bibliography{refs}

\begin{thebibliography}{23}
\providecommand{\natexlab}[1]{#1}
\providecommand{\url}[1]{\texttt{#1}}
\expandafter\ifx\csname urlstyle\endcsname\relax
  \providecommand{\doi}[1]{doi: #1}\else
  \providecommand{\doi}{doi: \begingroup \urlstyle{rm}\Url}\fi

\bibitem[Agrawal et~al.(2018)Agrawal, Verschueren, Diamond, and Boyd]{agrawal2018rewriting}
Akshay Agrawal, Robin Verschueren, Steven Diamond, and Stephen Boyd.
\newblock A rewriting system for convex optimization problems.
\newblock \emph{Journal of Control and Decision}, 5\penalty0 (1):\penalty0 42--60, 2018.

\bibitem[Boyd and Vandenberghe(2004)]{boyd2004convex}
Stephen Boyd and Lieven Vandenberghe.
\newblock \emph{Convex optimization}.
\newblock Cambridge university press, 2004.

\bibitem[Crippa and Orsini(2016)]{dosresmeta}
Alessio Crippa and Nicola Orsini.
\newblock Multivariate dose-response meta-analysis: The {dosresmeta} {R} package.
\newblock \emph{Journal of Statistical Software, Code Snippets}, 72\penalty0 (1):\penalty0 1--15, 2016.
\newblock \doi{10.18637/jss.v072.c01}.

\bibitem[Crippa et~al.(2019)Crippa, Discacciati, Bottai, Spiegelman, and Orsini]{crippa2019one}
Alessio Crippa, Andrea Discacciati, Matteo Bottai, Donna Spiegelman, and Nicola Orsini.
\newblock One-stage dose--response meta-analysis for aggregated data.
\newblock \emph{Statistical methods in medical research}, 28\penalty0 (5):\penalty0 1579--1596, 2019.

\bibitem[Dai et~al.(2022)Dai, Gil, Reitsma, Ahmad, Anderson, Bisignano, Carr, Feldman, Hay, He, et~al.]{dai2022health}
Xiaochen Dai, Gabriela~F Gil, Marissa~B Reitsma, Noah~S Ahmad, Jason~A Anderson, Catherine Bisignano, Sinclair Carr, Rachel Feldman, Simon~I Hay, Jiawei He, et~al.
\newblock Health effects associated with smoking: a burden of proof study.
\newblock \emph{Nature medicine}, 28\penalty0 (10):\penalty0 2045--2055, 2022.

\bibitem[Deeks et~al.(2019)Deeks, Higgins, Altman, and Group]{deeks2019analysing}
Jonathan~J Deeks, Julian~PT Higgins, Douglas~G Altman, and Cochrane Statistical~Methods Group.
\newblock Analysing data and undertaking meta-analyses.
\newblock \emph{Cochrane handbook for systematic reviews of interventions}, pages 241--284, 2019.

\bibitem[Diamond and Boyd(2016)]{diamond2016cvxpy}
Steven Diamond and Stephen Boyd.
\newblock {CVXPY}: {A} {P}ython-embedded modeling language for convex optimization.
\newblock \emph{Journal of Machine Learning Research}, 17\penalty0 (83):\penalty0 1--5, 2016.

\bibitem[Gautschi(1997)]{GautschiNumA}
Walter Gautschi.
\newblock \emph{Numerical analysis: an introduction}.
\newblock Birkhauser Boston Inc., USA, 1997.
\newblock ISBN 0817638954.

\bibitem[Grant et~al.(2006)Grant, Boyd, and Ye]{grant2006disciplined}
Michael Grant, Stephen Boyd, and Yinyu Ye.
\newblock \emph{Disciplined convex programming}.
\newblock Springer, 2006.

\bibitem[Greenland and Longnecker(1992)]{Greenland1992MethodsFT}
Sander Greenland and Matthew~P. Longnecker.
\newblock Methods for trend estimation from summarized dose-response data, with applications to meta-analysis.
\newblock \emph{American journal of epidemiology}, 135 11:\penalty0 1301--9, 1992.
\newblock URL \url{https://api.semanticscholar.org/CorpusID:31135711}.

\bibitem[Haidich(2010)]{haidich2010meta}
Anna-Bettina Haidich.
\newblock Meta-analysis in medical research.
\newblock \emph{Hippokratia}, 14\penalty0 (Suppl 1):\penalty0 29, 2010.

\bibitem[Hamling et~al.(2008)Hamling, Lee, Weitkunat, and Amb{\"u}hl]{hamling2008facilitating}
Jan Hamling, Peter Lee, Rolf Weitkunat, and Mathias Amb{\"u}hl.
\newblock Facilitating meta-analyses by deriving relative effect and precision estimates for alternative comparisons from a set of estimates presented by exposure level or disease category.
\newblock \emph{Statistics in medicine}, 27\penalty0 (7):\penalty0 954--970, 2008.

\bibitem[Itoga et~al.(2018)Itoga, Tawfik, Lee, Maruyama, Leeper, and Chang]{itoga2018association}
Nathan~K Itoga, Daniel~S Tawfik, Charles~K Lee, Satoshi Maruyama, Nicholas~J Leeper, and Tara~I Chang.
\newblock Association of blood pressure measurements with peripheral artery disease events: reanalysis of the allhat data.
\newblock \emph{Circulation}, 138\penalty0 (17):\penalty0 1805--1814, 2018.

\bibitem[Kariya and Kurata(2004)]{kariya2004generalized}
Takeaki Kariya and Hiroshi Kurata.
\newblock \emph{Generalized least squares}.
\newblock John Wiley \& Sons, 2004.

\bibitem[Lescinsky et~al.(2022)Lescinsky, Afshin, Ashbaugh, Bisignano, Brauer, Ferrara, Hay, He, Iannucci, Marczak, et~al.]{lescinsky2022health}
Haley Lescinsky, Ashkan Afshin, Charlie Ashbaugh, Catherine Bisignano, Michael Brauer, Giannina Ferrara, Simon~I Hay, Jiawei He, Vincent Iannucci, Laurie~B Marczak, et~al.
\newblock Health effects associated with consumption of unprocessed red meat: a burden of proof study.
\newblock \emph{Nature Medicine}, 28\penalty0 (10):\penalty0 2075--2082, 2022.

\bibitem[Liu et~al.(2009)Liu, Cook, Bergstr{\"o}m, and Hsieh]{liu2009two}
Qin Liu, Nancy~R Cook, Anna Bergstr{\"o}m, and Chung-Cheng Hsieh.
\newblock A two-stage hierarchical regression model for meta-analysis of epidemiologic nonlinear dose--response data.
\newblock \emph{Computational Statistics \& Data Analysis}, 53\penalty0 (12):\penalty0 4157--4167, 2009.

\bibitem[Orsini et~al.(2012)Orsini, Li, Wolk, Khudyakov, and Spiegelman]{orsini2012meta}
Nicola Orsini, Ruifeng Li, Alicja Wolk, Polyna Khudyakov, and Donna Spiegelman.
\newblock Meta-analysis for linear and nonlinear dose-response relations: examples, an evaluation of approximations, and software.
\newblock \emph{American journal of epidemiology}, 175\penalty0 (1):\penalty0 66--73, 2012.

\bibitem[Razo et~al.(2022)Razo, Welgan, Johnson, McLaughlin, Iannucci, Rodgers, Wang, LeGrand, Sorensen, He, et~al.]{razo2022effects}
Christian Razo, Catherine~A Welgan, Catherine~O Johnson, Susan~A McLaughlin, Vincent Iannucci, Anthony Rodgers, Nelson Wang, Kate~E LeGrand, Reed~JD Sorensen, Jiawei He, et~al.
\newblock Effects of elevated systolic blood pressure on ischemic heart disease: a burden of proof study.
\newblock \emph{Nature medicine}, 28\penalty0 (10):\penalty0 2056--2065, 2022.

\bibitem[Rudin et~al.(1964)]{rudin1964principles}
Walter Rudin et~al.
\newblock \emph{Principles of mathematical analysis}, volume~3.
\newblock McGraw-hill New York, 1964.

\bibitem[Schmidt and Kohlmann(2008)]{schmidt2008use}
Carsten~Oliver Schmidt and Thomas Kohlmann.
\newblock When to use the odds ratio or the relative risk?
\newblock \emph{International journal of public health}, 53\penalty0 (3):\penalty0 165, 2008.

\bibitem[Stanaway et~al.(2022)Stanaway, Afshin, Ashbaugh, Bisignano, Brauer, Ferrara, Garcia, Haile, Hay, He, et~al.]{stanaway2022health}
Jeffrey~D Stanaway, Ashkan Afshin, Charlie Ashbaugh, Catherine Bisignano, Michael Brauer, Giannina Ferrara, Vanessa Garcia, Demewoz Haile, Simon~I Hay, Jiawei He, et~al.
\newblock Health effects associated with vegetable consumption: a burden of proof study.
\newblock \emph{Nature medicine}, 28\penalty0 (10):\penalty0 2066--2074, 2022.

\bibitem[Zheng et~al.(2021)Zheng, Barber, Sorensen, Murray, and Aravkin]{zheng2021trimmed}
Peng Zheng, Ryan Barber, Reed~JD Sorensen, Christopher~JL Murray, and Aleksandr~Y Aravkin.
\newblock Trimmed constrained mixed effects models: formulations and algorithms.
\newblock \emph{Journal of Computational and Graphical Statistics}, 30\penalty0 (3):\penalty0 544--556, 2021.

\bibitem[Zheng et~al.(2022)Zheng, Aravkin, Murray, et~al.]{zheng2022burden}
Peng Zheng, Aleksandr Aravkin, Christopher Murray, et~al.
\newblock The burden of proof studies: assessing the evidence of risk.
\newblock \emph{Nature Medicine}, 28\penalty0 (10):\penalty0 2038--2044, 2022.

\end{thebibliography}
\pagebreak

\section{Appendix}\label{Appendix}
In this section, we provide proofs of theorems presented throughout this work.
\subsection{Proof of Theorem~\ref{thm:exist}}
    Take $G$ defined as in equation~\eqref{newt:obj}. $G$ is continuous on its domain $[0,\infty)^n$. First, we show that $G$ is proper, i.e., for some positive values of $A, a_0, B, b_0$, $G(A) \not \equiv +\infty$ and that for any $X\in[0,\infty)^n, G(X) > -\infty$. For this fact, we need the hypothesis in the statement of the theorem. Let $A$ is the vector of ones of length $n$, i.e., $A = [1,\dots, 1]^\top$. 
    Since, by hypothesis, $N_+ > A$ and $n_0 > a_0$, $G(A) < \infty$ inspection.   Also by inspection, $G$ is not equal to $-\infty$ for any $A$ in its domain.

    Next, $G$ is optimized over the compact set $0 \leq A \leq N$. 
   Since, by hypothesis, $N_+ > A$ and $n_0 > a_0$, $G(A) < \infty$ inspection.   Also by inspection, $G$ is not equal to $-\infty$ for any $A$ in its domain.  
   Since $G$ is continuous on the compact domain $0 \leq A \leq N$, it attains its minimum and maximum values.  Since $G$ is strictly convex, this minimizer must be unique.  This completes the proof.

\subsection{Proof of Theorem~\ref{thm:exist2}}
    Take $H$ as defined in equation~\eqref{obj:RR}. This proof will follow the same structure as the proof for Theorem~\ref{thm:exist}. We need only show $H$ is proper and that it has compact sublevel sets since $H$ is clearly continuous on the domain $[0,\infty )^n$. To show that $H$ is proper, similar to the proof of Theorem~\eqref{thm:exist}, consider the case when $A$ is the vector of ones of length $n$. By the hypothesis in the statement of Theorem~\ref{thm:exist2}, $n_0 > a_0$, so that $H$ is finite by inspection. Also by inspection, $H$ is never equal to $-\infty$ on any point in its domain. 

     Next, we show that $H$ has compact sublevel sets,
    that is,
    \[
    \mathcal{A}_\alpha := \{ A: H(A) \leq \alpha\}
    \]
    are closed and bounded. The closed prpoerty follows immediately by continuity. 
     Next, for a sequence of $X\in [0,\infty)^n$, $H(X)\to \infty$ as $\norm{X} \to\infty$ since $H$ is a sum of affine functions and entropic distance functions in all coordinates, see equation~\eqref{entropic_fn}. As $\norm{X} \to \infty$, the $x\log x$ terms in $H$ increase faster than linear terms. This implies directly that any sublevel set of $H$ must have an upper bound.  Thus, $H$ has compact sublevel sets. In particular, $H$ attains its minimum and maximum for any choice of sublevel set, so in particular we can consider $\alpha = H(1)$, the vector of all ones discussed in the previous paragraph. Once we know $H$ attains its minimum, we also know that the minimum is unique by strict convexity of the entropic distance.

\subsection{Proof of Theorem~\ref{thm:equivariant}}\label{pf:equivariant}
To prove the result, we simplify and rewrite the equations 
\[
\begin{aligned}
 \left(V- \frac{1}{a_0} - \frac{1}{b_0}\right)    \frac{1-p}{p}b_0 &= \sum_{i=1}^n \left(1+\frac{b_0}{a_0 R_i}\right) 
 = n + \frac{b_0}{a_0}\sum_{i=1}^n \frac{1}{R_i}
 = n + \frac{b_0}{a_0}r_1
 \\
   \left(V - \frac{1}{a_0} - \frac{1}{b_0}\right) \left(\frac{1}{zp} b_0 - a_0\right) &= \sum_{i=1}^n \left(1+\frac{a_0R_i}{b_0}\right)
   = n + \frac{a_0}{b_0} \sum_{i=1}^n R_i
 = n + \frac{a_0}{b_0}r_2
   \\
\end{aligned}
\]
Dividing the equations we obtain 
\[
 \frac{n + \frac{b_0}{a_0} r_1}{n + \frac{a_0}{b_0}r_2} = \frac{\frac{1-p}{p} b_0}{\frac{1}{zp} b_0 - a_0} 
 = 
 \frac{\frac{1-p}{p} }{\frac{1}{zp}  - \frac{a_0}{b_0}}
\]
Defining now $c = \frac{a_0}{b_0}$ we have 
\[
 \frac{n +  \frac{r_1}{c}}{n + cr_2}  =  \frac{\frac{1-p}{p} }{\frac{1}{zp}  - c}
\]
Multiplying by $c$ we have 
\[
 \frac{r_1 + nc}{n + r_2c}  =  \frac{(1-p)zc }{1  - pzc}
\]
The solution is given by 
\[
c = \frac{n p z - n z + n - p r_1 z \pm \sqrt{D}}{2 z (n p - p r_2 + r_2)}
\]
where 
\[
\begin{aligned}
D &= n^2 p^2 z^2 - 2 n^2 p z^2 + 2 n^2 p z + n^2 z^2 - 2 n^2 z + n^2 - 2 n p^2 r_1 z^2 + 2 n p r_1 z^2 + 2 n p r_1 z + p^2 r_1^2 z^2 - 4 p r_1 r_2 z + 4 r_1 r_2 z \\
& = n^2\left(p^2z^2-2pz^2+2pz + z^2-z+1\right) \\
& + n\left(-2p^2r_1z^2 + 2pr_1z^2 + 2pr_1z\right)\\
& + p^2r_1^2z^2 - 4pr_1r_2z + 4r_1r_2z
\end{aligned}
\]
We want to show that each piece is $\geq 0$. 
In fact we have 
\[
p^2z^2-2pz^2+2pz + z^2-z+1 = 
z\left((p-1)^2z + 2p -1\right) + 1
\]
The minimum with respect to $p$ of the inside expression occurs at $p-1 = \frac{-1}{z}$. Plugging in, that gives us 
\[
z\left(\frac{1}{z} + 1 - \frac{2}{z}\right) + 1 
= z,
\]
so as a result we have 
\[
n^2\left(p^2z^2-2pz^2+2pz + z^2-z+1\right) \geq n^2z. 
\]
Next, we have 
\[
n\left(-2p^2r_1z^2 + 2pr_1z^2 + 2pr_1z\right) 
= n(2r_1z)\left( (1-p)pz + p\right) \geq 2nr_1zp.
\]
Finally, we have 
\[
p^2r_1^2z^2 - 4pr_1r_2z + 4r_1r_2z 
= p^2r_1^2z^2 + 4(1-p)r_1r_2z \geq p^2r_1^2z^2
\]
Putting everything together, we get 
\[
D \geq n^2z + 2nr_1zp + p^2r_1^2z^2 
= z(n^2 + 2nr_1p + p^2r_1^2z) \geq 0.  
\]
Thus a solution always exists. 

To see that only one solution is positive, recall the form of the solution:
\[
c = \frac{n p z - n z + n - p r_1 z \pm \sqrt{D}}{2 z (n p - p r_2 + r_2)}
\]
We can observe that 
\[
 n^2\left(p^2z^2-2pz^2+2pz + z^2-z+1\right) - (n(z(p-1)+1))^2  = n^2z
\]
and as a result 
\[
\sqrt{D} - (n p z - n z + n - p r_1 z ) \geq 0.
\]
That means we have 
\[
c_2 = \frac{n p z - n z + n - p r_1 z - \sqrt{D}}{2 z (n p  +(1-p)  r_2 )} < 0 < c_1 = \frac{n p z - n z + n - p r_1 z + \sqrt{D}}{2 z (n p +(1- p) r_2)}. 
\]

Plugging $c_1$ in to the first equation, we have 
\[
b_0   = \frac{1}{V}\left(\frac{p}{1-p}\left(n + \frac{r_1}{c}\right) + 1 + \frac{1}{c_1}\right), \quad 
a_0 = c_1b_0. 
\]
and we have found the unique positive solution. This completes the proof.

\subsection{Proof of Theorem~\ref{thm:solution}}\label{pf:solution}
    We prove this theorem by induction. For the base case, when $n=1$, the existence of a unique positive solution follows immediately from Theorem ~\ref{thm:equivariant}.
    For the inductive hypothesis, suppose that for a given $n$ for the dimension of our vectors $V$ and $L$, we have the positive solution pair $a_0^n, b_0^n$ that simultaneously satisfy the system~\eqref{eq:nlEqs}. Thus, we have that
    \begin{align*}
        \frac{1-p}{p}b_0^n &= \sum_{i=1}^n \left(1+\frac{b_0^n}{a_0^n R_i}\right) / \left(V_i - \frac{1}{a_0^n} - \frac{1}{b_0^n}\right) \\
    \frac{1}{zp} b_0^n - a_0^n &= \sum_{i=1}^n \left(1+\frac{a_0^nR_i}{b_0^n}\right) / \left(V_i - \frac{1}{a_0^n} - \frac{1}{b_0^n}\right).
    \end{align*}
    If we continue to the step $n+1$, we  add strictly positive terms to the right hand side, and hence we have strict inequalities 
    \begin{align}\label{ineq:thm5}
    \begin{split}
        \frac{1-p}{p}b_0^n &< \sum_{i=1}^{n+1} \left(1+\frac{b_0^n}{a_0^n R_i}\right) / \left(V_i - \frac{1}{a_0^n} - \frac{1}{b_0^n}\right) \\
        \frac{1}{zp} b_0^n - a_0^n &< \sum_{i=1}^{n+1} \left(1+\frac{a_0^nR_i}{b_0^n}\right) / \left(V_i - \frac{1}{a_0^n} - \frac{1}{b_0^n}\right)
    \end{split}
    \end{align}
    and without loss of generality, we may assume that $V_{n+1} \geq \min_i V_i$  so that $V_{n+1} > \frac{1}{a_0^n} + \frac{1}{b_0^n}$. Otherwise, we can suitably reorder the terms and apply the inductive hypothesis.  
    
    Define the functions $f_1, f_2$ by
    \begin{align*}
        f_1(a_0,b_0) &= \frac{1-p}{p}b_0 - \sum_{i=1}^{n+1} \left(1+\frac{b_0}{a_0 R_i}\right) / \left(V_i - \frac{1}{a_0} - \frac{1}{b_0}\right) \\
        f_2(a_0,b_0) &= \frac{1}{zp} b_0 - a_0 - \sum_{i=1}^{n+1} \left(1+\frac{a_0R_i}{b_0}\right) / \left(V_i - \frac{1}{a_0} - \frac{1}{b_0}\right).
    \end{align*}
    
\begin{figure}[h!]
    \centering
    \begin{tikzpicture}
        \begin{axis}[
            xlabel={$f_1$},
            ylabel={$f_2$},
            xmin=-2.0, xmax=2.0,
            ymin=-1.5, ymax=1.5,
            axis lines=middle,
            grid=both,
            minor tick num=1,
            xticklabels={}, 
            yticklabels={} 
            ]
            \addplot[only marks] coordinates { (-1,-1) (-1,1) (1,-1) (1,1) (0,0)};
            \node[label={225:$(a_0^n,b_0^n)$},circle,fill,inner sep=2pt] at (axis cs:-1,-1) {};
            \node[label={135:$(a_0^3,b_0^3)$},circle,fill,inner sep=2pt] at (axis cs:-1,1) {};
            \node[label={315:$(a_0^1,b_0^1)$},circle,fill,inner sep=2pt] at (axis cs:1,-1) {};
            \node[label={45:$(a_0^2,b_0^2)$},circle,fill,inner sep=2pt] at (axis cs:1,1) {};
            \node[label={45:$(a_0^{n+1}, b_0^{n+1})$},circle,fill,inner sep=2pt] at (axis cs:0,0) {};
        \end{axis}
    \end{tikzpicture}
    \caption{ Note the generic form ($a_0,b_0$) is shorthand for $(f_1(a_0,b_0),f_2(a_0,b_0))$. The point $(a_0^{n+1}, b_0^{n+1})$ serves as desired solution point to complete the proof.
    \label{fig:points_plane}}
\end{figure}
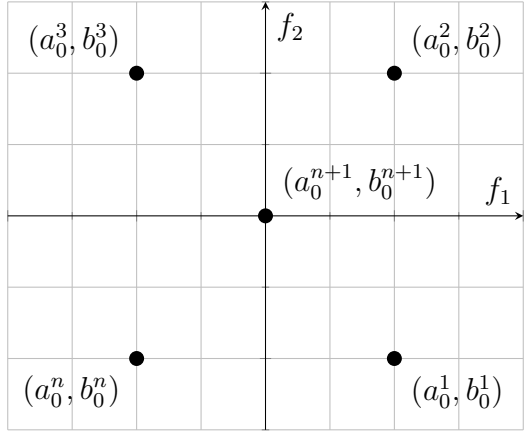

    The remaining work is focused on finding the $a_0^{n+1}, b_0^{n+1}$ such that $f_1(a_0^{n+1}, b_0^{n+1}) = f_2(a_0^{n+1}, b_0^{n+1}) = 0$, and is separated into two steps:
    \begin{enumerate}
        
        \item[Step 1] Show that we can find points  $(a_0^1, b_0^1)$, $(a_0^2, b_0^2)$, $(a_0^3, b_0^3)$ 
       with  
        \[
            f_1(a_0^1, b_0^1) > 0, \quad f_2(a_0^1, b_0^1) > 0,
        \]
        \[
        f_1(a_0^2,b_0^2) > 0, \quad f_2(a_0^2,b_0^2) < 0,
        \]
        and
        \[
            f_1(a_0^3,b_0^3) < 0, \quad f_2(a_0^3, b_0^3) > 0.
        \]
        These points, along with $(a_n, b_n)$ from the inductive hypothesis, are shown in Figure~\ref{fig:points_plane} and the four points set up continuation arguments used in Step 2. 
        
        \item[Step 2] Show that, by continuity of the solution maps, either case above leads to the existence of $(a_0^{n+1}, b_0^{n+1})$ simultaneously satisfying $f_1 = f_2 = 0$. 
    \end{enumerate}
    

\paragraph{Step 1 proof:}
  First, we observe that 
    \[
    \displaystyle \lim_{a_0 \uparrow \infty}f_1(a_0,b_0)  = 
    \frac{1-p}{p}b_0 - \sum_{i=1}^{n+1} \frac{1}{V_i - \frac{1}{b_0}}. 
    \]
    As long as we take $b_0^1 > 2\max\left(V, \frac{p}{1-p}2\sum_{i=1}^{n+1} \frac{1}{V_i}\right)$ where $V = \max_i V_i$, we can then find a large enough value $a_0^1$ satisfying $f_1(a_0^1, b_0^1) > 0$. 
    Next, we have 
    \[
    \lim_{a_0 \uparrow \infty}f_2(a_0,b_0) 
    = -\infty 
    \]
    for any $b_0$, so in particular we can select a large enough $a_0^1$ with $f_2(a_0^1,b_0^1) < 0$ and $f_1(a_0^1,b_0^1) > 0$. This gives us a point in the lower-right quadrant of Figure~\ref{fig:points_plane}. 
    
    Now we observe that 
    \[
    \lim_{b_0 \uparrow\infty}f_2(a_0, b_0) = \infty
    \]
    along the path $a_0 = \frac{1}{2} b_0$. 
    Along this same path, we have 
    \[
    \lim_{b_0 \uparrow \infty}f_1(a_0, b_0) = \infty
    \]
    as well.        
    We can thus select a large enough value  $b_0^2$
    and $a_0^2 = \frac{1}{2}b_0^2$ for which $f_1(a_0^2, b_0^2) > 0$ and $f_2(a_0^2, b_0^2) > 0$. This gives us the point in the upper-right quadrant of Figure~\ref{fig:points_plane}. 

    Next, consider $0 < \epsilon << 1$ and take 
    \[
    b_0= \frac{1}{\epsilon^2}, \quad a_0 = \frac{1}{V_{\min} - \epsilon - \epsilon^2}
    \]
    With these definitions, we have 
    \[
    f_2(\epsilon) \geq   \frac{1}{zp\epsilon^2} - \frac{1}{V_{\min} - \epsilon - \epsilon^2} - (n+1)\left(1 + \frac{\epsilon^2 R_i}{V_{\min} - \epsilon - \epsilon^2}\right)\frac{1}{\epsilon}  >0
    \]
\[
f_1(\epsilon) < \frac{1-p}{p\epsilon} - 
\sum_{i=1}^{n+1}\left(1 + \frac{V_{\min} - \epsilon - \epsilon^2}{\epsilon^2} \right)\frac{1}{\epsilon} < 0
\]
for small $\epsilon$. Thus for $0<\epsilon<<1$ we get a point $(a_0^3, b_0^3)$ with $f_2 > 0$ and $f_1 < 0$. This gives us a point in the upper left quadrant of Figure~\ref{fig:points_plane}.

Finally, by the inductive hypothesis, we have $f_1(a_0^n,b_0^n) < 0$ and $f_2(a_0^n,b_0^n) < 0$, 
which gives us a point in the lower left quadrant of Figure~\ref{fig:points_plane}.

\paragraph{Step 2 Proof:}
To show that there is a point $(a_0^{n+1},b_0^{n+1})$ such that the inequalities~\eqref{ineq:thm5} become equalities, we create separate interpolations 
    relying on the intermediate value theorem (IVT)~\cite{rudin1964principles}. 

    Consider the points $(a_0^n, b_0^n)$ and $(a_0^1, b_0^1)$ and the convex combination
    \[
    p_\lambda = \lambda(a_0^n, b_0^n) + (1-\lambda) (a_0^1, b_0^1)
    \]
    We have $f_1(p_1) < 0, f_1(p_0) > 0$, so by the IVT there is a $\lambda\in(0,1)$ with $f_1(p_\lambda) = 0$. 
    
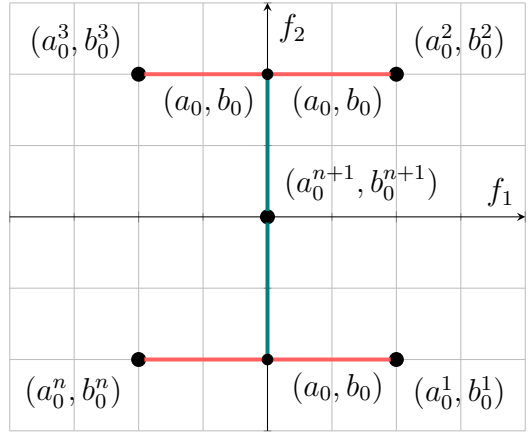
\begin{figure}[h!]
    \centering
    \begin{tikzpicture}
    \begin{axis}[
        xlabel={$f_1$},
        ylabel={$f_2$},
        xmin=-2.0, xmax=2.0,
        ymin=-1.5, ymax=1.5,
        axis lines=middle,
        grid=both,
        minor tick num=1,
        xticklabels={}, 
        yticklabels={}, 
    ]
        \addplot[only marks] coordinates { (-1,-1) (1,-1) (0,1) (0,-1) (0,0) (1,1) (-1,1)};
        \node[label={225:$(a_0^n,b_0^n)$},circle,fill,inner sep=2pt] at (axis cs:-1,-1) {};
        \node[label={315:$(a_0^1,b_0^1)$},circle,fill,inner sep=2pt] at (axis cs:1,-1) {};
        \node[label={45:$(a_0^{n+1}, b_0^{n+1})$},circle,fill,inner sep=2pt] at (axis cs:0,0) {};
        \node[label={45:$(a_0^2, b_0^2)$},circle,fill,inner sep=2pt] at (axis cs:1,1) {};
        \node[label={135:$(a_0^3, b_0^3)$},circle,fill,inner sep=2pt] at (axis cs:-1,1) {};
        \addplot[red!50!pink, line width=1.5pt] coordinates {(-1,-1) (1,-1)} node[black,anchor=north east]{$(a_0, b_0)$};
        \addplot[red!50!pink, line width=1.5pt] coordinates {(-1,1) (1,1)} node[black,anchor=north east]{$(a_0, b_0)$};
        \addplot[blue!50!green, line width=1.5pt] coordinates {(0,-1) (0,1)} node[black,anchor=north east]{$(a_0, b_0)$};
    \end{axis}
\end{tikzpicture}
    \caption{The points and their associated continuous deformations (the colored lines) according to Case 1. Note the generic form ($a_0,b_0$) is shorthand for $(f_1(a_0,b_0),f_2(a_0,b_0))$. 
    \label{fig:cont_deform1}}
\end{figure}

    If $f_2(p_\lambda) > 0$, 
    we proceed to Case 1 below. 
    If $f_2(p_\lambda) < 0$, 
    we have a point of intersection below the $f_1$ axis, as shown in Figure~\ref{fig:cont_deform1}. 
    We then apply IVT to $(a_0^3, b_0^3)$ and $(a_0^2, b_0^2)$. If the crossing point obtained from the IVT is above the $f_1$ axis, we proceed to Case 2 below, and if it is below the $f_1$ axis, we proceed to Case 1 below.      
    
\begin{figure}[h!]
    \centering
    \begin{tikzpicture}
    \begin{axis}[
        xlabel={$f_1$},
        ylabel={$f_2$},
        xmin=-2.0, xmax=2.0,
        ymin=-1.5, ymax=1.5,
        axis lines=middle,
        grid=both,
        minor tick num=1,
        xticklabels={}, 
        yticklabels={}, 
    ]
        \addplot[only marks] coordinates { (-1,-1) (1,-1) (0,0) (-.577,0) (.577,0) (1,1) (-1,1)};
        \node[label={225:$(a_0^n,b_0^n)$}, circle, fill, inner sep=2pt] at (axis cs:-1,-1) {};
        \node[label={315:$(a_0^1, b_0^1)$}, circle, fill, inner sep=2pt] at (axis cs:1,-1) {};
        \node[label={270:$(a_0^{n+1},b_0^{n+1})$}, circle, fill, inner sep=2pt] at (axis cs:0,0) {};
        \node[label={45:$(a_0^2, b_0^2)$}, circle, fill, inner sep=2pt] at (axis cs:1,1) {};
        \node[label={135:$(a_0^3, b_0^3)$}, circle, fill, inner sep=2pt] at (axis cs:-1,1) {};
        \addplot[red!50!pink, line width=1.5pt] coordinates {(1,1) (-1,1)};
        \addplot[blue!50!pink, line width=1.5pt] coordinates {(-.577,0) (.577,0)};
        \addplot[blue, domain=-1:1, samples=100, line width=1.5pt] {-1.5*(x)^2 + 0.5};
    \end{axis}
\end{tikzpicture}
    \caption{The points and their associated continuous deformations (the colored lines) according to Case 2 in the $f_1-f_2$ plane. Note the generic form $(a_0,b_0)$ is shorthand for $(f_1(a_0,b_0),f_2(a_0,b_0))$. 
    \label{fig:cont_deform2}}
\end{figure}
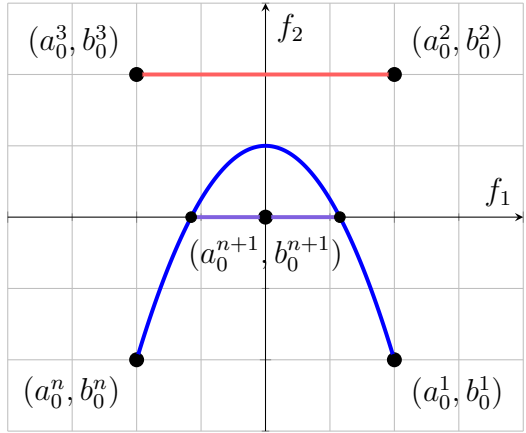

    \begin{itemize}
\item[Case 1:] $f_2(p_\lambda) > 0$ or IVT applied to $(a_0^3, b_0^3)$ and $(a_0^2, b_0^2)$ yields a point below the $f_1$ axis. In either case, applying the IVT twice, we obtain two points along the $f_1$ axis, as shown in Figure~\ref{fig:cont_deform2},   
with opposite signs along $f_1$. The constraint $f_2 = 0$ is easily incorporated into $f_1$, which becomes 
\begin{align*}
f_3(a_0, b_0) &= (1-p)z\left(a_0 + \sum_{i=1}^{n+1} \left(1+\frac{a_0R_i}{b_0}\right) / \left(V_i - \frac{1}{a_0} - \frac{1}{b_0}\right)\right) \\
&- \sum_{i=1}^{n+1} \left(1+\frac{a_0R_i}{b_0}\right) / \left(V_i - \frac{1}{a_0} - \frac{1}{b_0}\right)
\end{align*}
Clearly $f_3$ has opposite signs for the two points of intersection in Figure~\ref{fig:cont_deform2}, and applying IVT again we find the point $(a_0^{n+1}, b_0^{n+1})$. 

\item[Case 3:] In this case, we have successfully found two points with $f_1 = 0$, and opposite signs with respect to $f_2$. Just as in the previous case, we can explicitly incorporate the constraint $f_1 = 0$ into $f_2$, to obtain 
\begin{align*}
f_4(a_0, b_0) &= \frac{1}{zp(1-p)}  
 \sum_{i=1}^{n+1} \left(1+\frac{b_0}{a_0 R_i}\right) / \left(V_i - \frac{1}{a_0} - \frac{1}{b_0}\right)\\
&- a_0 - \sum_{i=1}^{n+1} \left(1+\frac{a_0R_i}{b_0}\right) / \left(V_i - \frac{1}{a_0} - \frac{1}{b_0}\right)
\end{align*}
Clearly $f_4$ has opposite signs now for the two crossing points shown on the $f_2$ axis of Figure~\ref{fig:cont_deform1}, and by IVT, we have existence of $(a_0^{n+1}, b_0^{n+1})$. 
    \end{itemize}

\subsection{Proof of Theorem~\ref{thm:equivariantRR}}\label{pf:equivariantRR}

To prove the result, we simplify and rewrite the equations 
\[
\begin{aligned}
\left(V_i - \frac{1}{a_0} + \frac{1}{b_0}\right)   \frac{1-p}{p}b_0 &= \sum_{i=1}^n \left(\frac{b_0}{a_0 R_i} - 1\right)
 = \frac{b_0}{a_0}\sum_{i=1}^n \frac{1}{R_i}-n
 = \frac{b_0}{a_0}r_1-n
 \\
  \left(V_i - \frac{1}{a_0} + \frac{1}{b_0}\right)
  \left(\frac{1}{zp} b_0 - a_0\right) &= \sum_{i=1}^n \left(1-\frac{a_0R_i}{b_0}\right) 
   = n - \frac{a_0}{b_0} \sum_{i=1}^n R_i
 = n - \frac{a_0}{b_0}r_2
\end{aligned}
\]
Dividing the equations we obtain 
\[
 \frac{\frac{b_0}{a_0} r_1-n}{n - \frac{a_0}{b_0}r_2} = \frac{\frac{1-p}{p} b_0}{\frac{1}{zp} b_0 - a_0} 
 = 
 \frac{\frac{1-p}{p} }{\frac{1}{zp}  - \frac{a_0}{b_0}}
\]
Defining now $c = \frac{a_0}{b_0}$ we have 
\[
 \frac{\frac{r_1}{c}-n}{n - cr_2}  =  \frac{\frac{1-p}{p} }{\frac{1}{zp}  - c}
\]
with the inherited constraint that $c < 1$, since $a_0 < b_0$ by definition. Multiplying by $c$ we have 
\[
 \frac{r_1 - nc}{n - r_2c}  =  \frac{(1-p)zc }{1  - pzc}
\]
The solution is given by 
\[
c = \frac{n(z-pz + 1) + p r_1 z \pm \sqrt{D}}{2 z (n p +(1-p) r_2)}
\]
where 
\[
D = (n(pz-z-1)-r_1zp)^2-4r_1(nzp+r_2z-r_2pz). 
\]
By inspection the coefficient for $n^2$ is given by 
\[
(pz-z-1)^2, 
\]
which is always positive. The coefficient for $n$ is given by 
\[
(-2(pz-z-1)-4)zpr_1 = 2zpr_1(z-pz-1)
\]
This term is non-negative exactly when $(1-p)z \geq 1$.
Finally, the constant term is given by 
\[
r_1^2z^2p^2+4r_1r_2z(p-1).
\]
The condition for when this term is non-negative can also be written in terms of $(1-p)z$:
\[
(1-p)z \geq \left(\frac{1-p}{p}\right)^24r_2.
\]
It is easy to find a counter-example when these conditions are violated, and where $D<0$. 
\[
n  = 2, \quad p = 0.1, \quad z = 1.1, \quad  r_1 = 31.9,  \quad r_2 = 1. 
\]
This yields $D = -33.4$, so there is no solution. 
In this case, $(1-p)z = 0.99$, while 
$\frac{(1-p)^2}{p^2}4r_2 = 324$, so both inequalities are violated, the latter significantly. 
The result is easily achievable, 
since we just want to find $\alpha$ with 
\[
\frac{2}{1-\alpha} + \frac{2}{\alpha} = 31.9.
\]
This gives us
\[
R_1 = .9328, R_2 = .0672.
\]

\end{document}